\documentclass[12pt]{article} 
\pdfoutput=1
\usepackage{graphicx,floatflt,amssymb,rotate}
\usepackage{mathrsfs}
\usepackage{amssymb}
\usepackage{amsfonts}
\usepackage{amsmath}
\usepackage{color}
\usepackage{graphicx}
\usepackage{subfig}
\usepackage{multirow}
\usepackage{float}
\usepackage{pstricks}
\usepackage[toc,page]{appendix}
\usepackage{cite}
\usepackage{soul}
\usepackage[colorlinks=true,
linkcolor=red,
urlcolor=blue,
citecolor=blue]{hyperref}
\usepackage{color}

\newcommand{\vcb}{|V_{cb}|}
\newcommand{\vtd}{|V_{td}|}
\newcommand{\vub}{|V_{ub}/V_{cb}|}
\newcommand{\vts}{|V_{ts}|}

\def\R1{\varepsilon_1}
\def\E8{\varepsilon_8}

\newcommand{\mt}{m_{\rm t}}

\newcommand{\bea}{\begin{eqnarray}}
\newcommand{\eea}{\end{eqnarray}}
\newcommand{\bd}{\begin{displaymath}}
\newcommand{\ed}{\end{displaymath}}

\newcommand{\be}{\begin{equation}}
\newcommand{\ee}{\end{equation}}
\newcommand{\bi}{\begin{itemize}}
\newcommand{\ei}{\end{itemize}}
\newcommand{\ord}{{\cal O}}

\begin{document}
\vskip 30pt  
 
\begin{center}  
{\Large \bf The impact of nonminimal Universal Extra Dimensional model on \boldmath$\Delta B=2$ transitions} \\
\vspace*{1cm}  
\renewcommand{\thefootnote}{\fnsymbol{footnote}}  
{{\sf Avirup Shaw$^1$\footnote{email: avirup.cu@gmail.com}} 
}\\  
\vspace{10pt}  
{{\em $^1$Theoretical Physics, Physical Research Laboratory,\\
Ahmedabad 380009, India}}
\normalsize  
\end{center} 

\begin{abstract}
\noindent 
{We measure the impact of nonvanishing boundary localised terms on $\Delta B=2$ transitions in five-dimensional Universal Extra Dimensional scenario where masses and coupling strengths of several interactions of Kaluza-Klein modes are significantly modified with respect to the minimal counterpart. In such scenario we estimate the Kaluza-Klein contributions of quarks, gauge bosons and charged Higgs by evaluating the one-loop box diagrams that are responsible for the $\Delta B=2$ transitions. Using the loop function (obtained from one-loop box diagrams) we determine several important elements that are involved in Wolfenstein parametrisation. Moreover, with these elements we also study the geometrical shape of unitarity triangle. Besides, we compute the quantity $\Delta M_s$ scaled by the corresponding Standard Model value. Outcomes of our theoretical predictions have been compared to the allowed ranges of the corresponding observables simultaneously. Our current analysis shows that, depending on the parameters in this scenario the lower limit on the inverse of the radius of compactification can reach to an appreciable large value ($\approx 1.48$ TeV or even higher).
}
\vskip 5pt \noindent 
\end{abstract}

\renewcommand{\thesection}{\Roman{section}}  
\setcounter{footnote}{0}  
\renewcommand{\thefootnote}{\arabic{footnote}}

\section{Introduction}
{The Standard Model (SM) of particle physics has been a tremendously successful theory for explaining the features and interactions of fundamental particles, with many measurements confirming its predictions to extraordinary precision. With the discovery of the Higgs Boson by the ATLAS\cite{Aad:2012tfa} and CMS\cite{Chatrchyan:2012xdj} at the Large Hadron Collider (LHC) at CERN, all the particles in the SM zoo have been observed experimentally. However, it is known to be incomplete, as there exist several experimental data, such as massive neutrinos, Dark Matter (DM), matter anti-matter asymmetry etc., that cannot be explained in the SM scenario. Therefore, one of the current goals of particle physics is to discover new particles and interactions-generically known as ``new physics" (NP) that could provide an explanation for these observations. In principle, there are two ways to search for NP. At the high-energy frontier one tries to produce those new degrees of freedom directly, while at the high-precision frontier one analyses the indirect virtual effects of such new particles. In the second method NP would appear as a discrepancy between SM expectations and experimental measurements. 

Considering the latter argument, we would like to mention that, one of the elegant ways to search for new particles is by studying processes known as flavour changing neutral current (FCNC) decays, where a quark changes its flavour without changing its electric charge. One example of such a transition is the decay of a bottom quark ($b$) into a strange ($s$) or down ($d$) quark. In such classification, $B_q$-{meson} $(q=s, d)$ mixing is a particularly interesting process for indirect NP searches in the quark-flavour sector. Since in the SM such processes are forbidden at tree-level, they are sensitive to new heavy particles appearing as virtual particles in loop diagrams. Moreover, it is Glashow Iliopoulos Maiani (GIM) suppressed. The physical observables are the mass differences ($\Delta M_q$), decay-width differences ($\Delta \Gamma_q$) between the heavy and light neutral $B_q$-meson mass eigenstates, and the flavour-specific CP asymmetries ($a^q_{\rm fs}$). Theoretical predictions of $B_q$-mixing observables in both the SM and beyond are governed by $\Delta B=2$ transitions and the hadronic matrix elements of which are expressed by local four-fermion operators in the effective weak Hamiltonian (given in Eq.\;\ref{hdb2}). These observables are very useful to constrain physics beyond SM (BSM) scenarios. For example, in any BSM scenario if we compute the contribution of new heavy particles (particularly the contribution in box diagram for $B_q$-mixing) then it will contribute to loop-integral function ( Inami-Lim function \cite{10.1143/PTP.65.297}) of the corresponding operator which controls the $\Delta B=2$ transitions. Consequently, using the $B_q$-mixing observables we can measure the effects of that BSM scenarios by extracting the elements of the Cabibbo Kobayashi Maskawa (CKM) matrix \cite{PhysRevLett.10.531, 10.1143/PTP.49.652} as well as by studying the shape of unitarity triangle (UT) \cite{Buras:2000dm}. Generally, four independent parameters are required to define the CKM matrix fully. Out of many parametrisations, Wolfenstein parametrisation \cite{Wolfenstein:1983yz} is the most famous and has several nice features. In particular it gives very prominent geometrical representation of the structure of the CKM matrix in conjunction with the UT which in turn very helpful to constrain the BSM scenarios. For example the refs. \cite{Chakraverty:2002qk, Buras:2002ej, Choudhury:2004bh, Altmannshofer:2007cs, Carlucci:2009gr, Hu:2019heu} depicted how the observables are related to $\Delta B=2$ transition that can constrain different BSM scenarios. In the current article we consider a class of models, namely Universal Extra Dimensional (UED) \cite{Appelquist:2000nn} scenario, with {\it nonvanishing boundary localised terms (BLTs)}, where the low energy effective Hamiltonians are controlled by local operators that are same as in the SM. In this type of scenario flavour violation and CP violation are entirely governed by the CKM matrix, with Minimal Flavour Violation (MFV) as defined in \cite{Buras:2000dm, Buras:2002yj, DAmbrosio:2002vsn, Buras:2002ej}.}

{

UED \cite{Appelquist:2000nn} scenario is a specific extension of SM with one flat space-like dimension ($y$) compactified on a circle $S^1$ of radius $R$. Each of the SM fields is exposed to the extra dimension $y$. The fields appeared on this manifold are generally defined as towers of 4-dimensional (4D) Kaluza-Klein (KK) states while the zero-mode of the KK-towers is recognised as the corresponding 4D SM field. SM chiral fermions are emerged in this scenario by imposing a discrete symmetry ${Z}_2$ ($y \leftrightarrow -y$) on the extra spatial dimension. Therefore, the extra dimension is known as an $S^1/Z_2$ orbifold and as a consequence physical domain extends from $y = 0$ to $y = \pi R$. Eventually, the $y \leftrightarrow -y$ symmetry is appeared as a conserved parity which is designated as KK-parity  $=(-1)^n$, where $n$ is known as KK-number and it measures the discretised momentum along the $y$-direction. Due to the conservation of KK-parity the lightest Kaluza-Klein particle (LKP) with KK-number one ($n=1$) becomes absolutely stable and cannot decay to a pair of SM particles. Therefore, the LKP has been treated as a potential DM candidate in this scenario \cite{Servant:2002hb, Servant:2002aq, Cheng:2002ej, Majumdar:2002mw, Burnell:2005hm, Kong:2005hn, Kakizaki:2006dz, Belanger:2010yx}. Besides, a few alternatives of this model can resolve some other demerits of SM, for example, gauge coupling unifications \cite{Dienes:1998vh, Dienes:1998vg, Bhattacharyya:2006ym}, neutrino mass \cite{Hsieh:2006qe, Fujimoto:2014fka} and fermion mass hierarchy \cite{Archer:2012qa} etc. 

At the $n^{th}$ KK-level the mass of KK-partner of any SM particle can be expressed as $\sqrt{(m^2+(nR^{-1})^2)}$\;, where $m$ is identified as the zero-mode mass (SM particle mass) and it is very small in comparison to $R^{-1}$. Eventually, this UED scenario consists of nearly degenerate mass spectrum at each KK-level. Due to this reason, UED scenario suffers from lack of phenomenological importance, particularly, at the colliders. However, radiative corrections \cite{Georgi:2000ks, Cheng:2002iz} can cure the problem of degeneracy in the mass spectrum. The radiative corrections can be divided into two categories, e.g., the first one is bulk corrections (which are finite and only nonzero for KK-excitations of gauge bosons) while the other one is considered as boundary localised corrections. The latter is proportional to logarithmically cut-off\footnote{UED is regarded as an effective theory and it is characterised by a cut-off scale $\Lambda$.} scale ($\Lambda$) dependent terms. One can allow the boundary correction terms as 4D mass, kinetic and other possible interaction terms for the KK-excited states at the two fixed boundary points ($y=0$ and $y=\pi R$) of this orbifold. Actually, it is very natural to consider such terms in an extra dimensional theory like UED, because these boundary terms have served as the counterterms for cut-off dependent loop-induced contributions. There is a special assumption in the minimal version of UED (mUED) model, where the boundary terms are chosen in such a way that the 5D radiative corrections are disappeared at the cut-off scale $\Lambda$. Although, this unique assumption can be discarded and without computing the exact radiative corrections one could parametrise these as kinetic, mass as well as other interaction terms localised at the two fixed boundary points. Hence, this typical version is known as nonminimal Universal Extra Dimensional (nmUED) model \cite{Dvali:2001gm, Carena:2002me, delAguila:2003bh, delAguila:2003kd, delAguila:2003gv, Schwinn:2004xa, Flacke:2008ne, Datta:2012xy, Flacke:2013pla}. Within this scenario, apart from the radius of compactification ($R$), coefficients of different BLTs have been considered as free parameters and that can be constrained by various experimental data of several physical observables. In literature there exists number of phenomenological studies in this scenario. For example, bounds on the values of the coefficients of the BLTs have been obtained from the evaluation of electroweak observables \cite{Flacke:2008ne, Flacke:2013pla}, S, T and U parameters \cite{delAguila:2003gv, Flacke:2013nta}, DM relic density \cite{Bonnevier:2011km, Datta:2013nua}, production as well as decay of SM Higgs boson \cite{Dey:2013cqa}, collider study of LHC experiments \cite{Datta:2012tv, Datta:2013yaa, Datta:2013lja, Shaw:2014gba, Shaw:2017whr, Ganguly:2018pzs}, $Z\to b\bar{b}$ \cite{Jha:2014faa}, branching ratios of some rare decay processes of $B$-meson: e.g., $B_s \rightarrow \mu^+ \mu^-$ \cite{Datta:2015aka}, $B \to X_s\gamma$ \cite{Datta:2016flx} and $B \to X_s\ell^+\ell^-$\cite{Shaw:2019fin}, $\mathcal{R}_{D^{(*)}}$ anomalies \cite{Biswas:2017vhc, Dasgupta:2018nzt, Lee:2019phf}, flavour changing rare top decay \cite{Dey:2016cve, Chiang:2018oyd} and unitarity of scattering amplitudes containing KK-excitations \cite{Jha:2016sre}. 

{Within the scope of this article, to the best of our knowledge, for the first time we explore the $\Delta B=2$ transitions in the nmUED scenario by computing the KK-contributions to the one-loop box diagrams. The function emerging from the box diagrams is not only affected by the radius of compactification but also by the BLT parameters. Now it has already been mentioned that with the help of this function we can extract the CKM elements which in turn gives geometrical representation of the UT. Therefore, if we compare our theoretical prediction with the current allowed ranges of the CKM parameters and elements of the UT then we can easily constrain the parameter space of this nmUED framework. Moreover, from our study we can also measure the lower limit on $R^{-1}$ and compare the same with the results obtained from our previous analyses \cite{Datta:2015aka, Datta:2016flx, Shaw:2019fin}. Similar kind of exercise was executed several years ago in UED framework \cite{Buras:2002ej} where the BLT parameters are zero. In the present article, considering the current allowed ranges of the observables \cite{Zyla:2020zbs, CKM:summer} we will also revisit the lower bound on {$R^{-1}$} in UED framework with the BLT parameters as zero.} 

The paper is organised as follows. We will give a brief description of the nmUED model in section \ref{model}. Then in section \ref{bbbar_mixing} we will present the calculational details of $\Delta B=2$ transition in nmUED scenario. Consequently, we show the mechanism of the extraction of CKM parameters and the elements of UT. In section \ref{anls} we will present our numerical results. Finally, we will summarise the results in section \ref{concl}.

\section{A concise overview of KK-parity conserving nmUED scenario}\label{model}
In this section we overview the salient features of the nmUED scenario necessary for our current analysis. One can find detailed description of this scenario in\cite{Dvali:2001gm, Carena:2002me, delAguila:2003bh, delAguila:2003kd, delAguila:2003gv, Schwinn:2004xa, Flacke:2008ne, Datta:2012xy, Datta:2012tv, Datta:2013yaa, Datta:2013lja, Shaw:2014gba, Shaw:2017whr, Ganguly:2018pzs, Jha:2014faa, Datta:2015aka, Datta:2016flx, Shaw:2019fin, Biswas:2017vhc}. In order to conserve the KK-parity, coefficients of boundary terms at both the boundary points ($y=0$ and $y=\pi R$) are kept equal. Therefore, one has a stable LKP in this scenario and hence the present scenario can provide a potential DM candidate (such as first excited KK-state of the photon). In this type of scenario, one can find an extensive study on DM in \cite{Datta:2013nua}.
 
Action for 5D fermionic fields considering appropriate boundary localised kinetic term (BLKT) with coefficient $r_f$ \cite{Schwinn:2004xa, Datta:2013nua, Datta:2015aka, Datta:2016flx, Shaw:2019fin, Biswas:2017vhc} can be written as 
\begin{eqnarray} 
S_{fermion} = \int d^5x \left[ \bar{\Psi}_L i \Gamma^M D_M \Psi_L 
+ r_f\{\delta(y)+\delta(y - \pi R)\} \bar{\Psi}_L i \gamma^\mu D_\mu P_L\Psi_L  
\right. \nonumber \\
\left. + \bar{\Psi}_R i \Gamma^M D_M \Psi_R
+ r_f\{\delta(y)+\delta(y - \pi R)\}\bar{\Psi}_R i \gamma^\mu D_\mu P_R\Psi_R
\right]\;.
\label{factn}
\end{eqnarray}
In the above action 5D four component Dirac spinors are represented by $\Psi_L(x,y)$ and $\Psi_R(x,y)$, which can be expressed in terms of two component spinors as \cite{Schwinn:2004xa, Datta:2013nua, Datta:2015aka, Datta:2016flx, Shaw:2019fin, Biswas:2017vhc}
\begin{equation} 
\Psi_L(x,y) = \begin{pmatrix}\phi_L(x,y) \\ \chi_L(x,y)\end{pmatrix}
=   \sum_n \begin{pmatrix}\phi^{(n)}_L(x) f_L^n(y) \\ \chi^{(n)}_L(x) g_L^n(y)\end{pmatrix}, 
\label{fermionexpnsn1}
\end{equation}
\begin{equation} 
\Psi_R(x,y) = \begin{pmatrix}\phi_R(x,y) \\ \chi_R(x,y) \end{pmatrix} 
=   \sum_n \begin{pmatrix}\phi^{(n)}_R(x) f_R^n(y) \\ \chi^{(n)}_R(x) g_R^n(y) \end{pmatrix}. 
\label{fermionexpnsn2} 
\end{equation}\\
Here, $f_{L(R)}$ and $g_{L(R)}$ are the KK-wave-functions that can be shown as the following form \cite{Carena:2002me, Flacke:2008ne, Datta:2013nua, Datta:2015aka, Datta:2016flx, Shaw:2019fin, Biswas:2017vhc}
\begin{eqnarray}
f_L^n = g_R^n = N^f_n \left\{ \begin{array}{rl}
                \displaystyle \frac{\cos\left[m_{f^{(n)}} \left (y - \frac{\pi R}{2}\right)\right]}{\cos[ \frac{m_{f^{(n)}} \pi R}{2}]}  &\mbox{for $n$ even,}\\
                \displaystyle \frac{{-}\sin\left[m_{f^{(n)}} \left (y - \frac{\pi R}{2}\right)\right]}{\sin[ \frac{m_{f^{(n)}} \pi R}{2}]} &\mbox{for $n$ odd,}
                \end{array} \right.
                \label{flgr}
\end{eqnarray}
and
\begin{eqnarray}
g_L^n =-f_R^n = N^f_n \left\{ \begin{array}{rl}
                \displaystyle \frac{\sin\left[m_{f^{(n)}} \left (y - \frac{\pi R}{2}\right)\right]}{\cos[ \frac{m_{f^{(n)}} \pi R}{2}]}  &\mbox{for $n$ even,}\\
                \displaystyle \frac{\cos\left[m_{f^{(n)}} \left (y - \frac{\pi R}{2}\right)\right]}{\sin[ \frac{m_{f^{(n)}} \pi R}{2}]} &\mbox{for $n$ odd.}
                \end{array} \right.
\end{eqnarray}
In the above expressions, $N^f_n$ represents the normalisation constant for $n^{th}$ KK-mode wave-function and can readily be derived from the following orthonormality conditions \cite{Datta:2013nua, Datta:2015aka, Datta:2016flx, Shaw:2019fin, Biswas:2017vhc}
\begin{equation}\label{orthonorm}
\begin{aligned}
&\left.\begin{array}{r}
                  \int_0 ^{\pi R}
dy \; \left[1 + r_{f}\{ \delta(y) + \delta(y - \pi R)\}\right]f_L^mf_L^n\\
                  \int_0 ^{\pi R}
dy \; \left[1 + r_{f}\{ \delta(y) + \delta(y - \pi R)\}\right]g_R^mg_R^n
\end{array}\right\}=&&\delta^{n m}~;
&&\left.\begin{array}{l}
                 \int_0 ^{\pi R}
dy \; f_R^mf_R^n\\
                 \int_0 ^{\pi R}
dy \; g_L^mg_L^n
\end{array}\right\}=&&\delta^{n m}~,
\end{aligned}
\end{equation}
and the compact form of this normalisation constant is given by \cite{Datta:2013nua, Datta:2015aka, Datta:2016flx, Shaw:2019fin, Biswas:2017vhc}
\begin{equation}\label{norm}
N^f_n=\sqrt{\frac{2}{\pi R}}\Bigg[ \frac{1}{\sqrt{1 + \frac{r^2_f m^2_{f^{(n)}}}{4} + \frac{r_f}{\pi R}}}\Bigg].
\end{equation}

KK-mass of $n^{th}$ KK-excitation is represented by $m_{f^{(n)}}$ and it can be obtained from the following transcendental equations \cite{Carena:2002me, Datta:2013nua, Datta:2015aka, Datta:2016flx, Shaw:2019fin, Biswas:2017vhc} 
\begin{eqnarray}
  \frac{r_{f} m_{f^{(n)}}}{2}= \left\{ \begin{array}{rl}
         -\tan \left(\frac{m_{f^{(n)}}\pi R}{2}\right) &\mbox{for $n$ even,}\\
          \cot \left(\frac{m_{f^{(n)}}\pi R}{2}\right) &\mbox{for $n$ odd.}
          \end{array} \right.   
          \label{fermion_mass}      
 \end{eqnarray}
To this end, we would like to discuss the Yukawa interactions in this scenario, as the large top quark mass plays a pivotal role in enhancing the quantum effects in the present work. The action of Yukawa interaction including BLTs with coefficient $r_y$ is given by \cite{Datta:2015aka, Datta:2016flx, Shaw:2019fin, Biswas:2017vhc}  
\begin{eqnarray}
\label{yukawa}
S_{Yukawa} &=& -\int d^5 x  \Big[\lambda^5_t\;\bar{\Psi}_L\widetilde{\Phi}\Psi_R 
  +r_y \;\{ \delta(y) + \delta(y-\pi R) \}\lambda^5_t\bar{\phi_L}\widetilde{\Phi}\chi_R+\textrm{h.c.}\Big].
\end{eqnarray}
In the above action $\lambda^5_t$ represents the 5D coupling strength of Yukawa interaction for the third generations. $\Phi=\left(\begin{array}{cc} \phi^+\\\phi^0\end{array}\right)$ is the 5D Higgs doublet field and $\widetilde{\Phi}=i\sigma^2\Phi^*$. Inserting the KK-wave-functions for fermions (given in Eqs.\;\ref{fermionexpnsn1} and \ref{fermionexpnsn2}) in the actions given in Eq.\;\ref{factn} and Eq.\;\ref{yukawa}, one obtains the bi-linear terms containing the doublet and singlet states of the quarks. The resulting mass matrix for the $n^{th}$ KK-level for third generation of quark can be written as the following \cite{Datta:2015aka, Datta:2016flx, Shaw:2019fin, Biswas:2017vhc}
\begin{equation}
\label{fermion_mix}
-\begin{pmatrix}
\bar{\phi_L}^{(n)} & \bar{\phi_R}^{(n)}
\end{pmatrix}
\begin{pmatrix}
m_{f^{(n)}}\delta^{nm} & m_{t} {\mathscr{I}}^{nm}_1 \\ m_{t} {\mathscr{I}}^{mn}_2& -m_{f^{(n)}}\delta^{mn}
\end{pmatrix}
\begin{pmatrix}
\chi^{(m)}_L \\ \chi^{(m)}_R
\end{pmatrix}+{\rm h.c.}\;,
\end{equation}
where $m_t$ represents the mass of SM top quark and $m_{f^{(n)}}$ is derived from the solution of the transcendental equations given in Eq.\;\ref{fermion_mass}. ${\mathscr{I}}^{nm}_1$ and ${\mathscr{I}}^{nm}_2$ are the overlap integrals that are given in the following\cite{Datta:2015aka, Datta:2016flx, Shaw:2019fin, Biswas:2017vhc}
  \[ {\mathscr{I}}^{nm}_1=\left(\frac{1+\frac{r_f}{\pi R}}{1+\frac{r_y}{\pi R}}\right)\times\int_0 ^{\pi R}\;dy\;
\left[ 1+ r_y \{\delta(y) + \delta(y - \pi R)\} \right] g_{R}^m f_{L}^n,\] \;\;{\rm and}\;\;\[{\mathscr{I}}^{nm}_2=\left(\frac{1+\frac{r_f}{\pi R}}{1+\frac{r_y}{\pi R}}\right)\times\int_0 ^{\pi R}\;dy\;
 g_{L}^m f_{R}^n .\]

The integral ${\mathscr{I}}^{nm}_1$ is nonvanishing for both the cases of $n=m$ and $n\neq m$. However, in the limit $r_y = r_f$, this integral becomes unity (when $n =m$) or zero ($n \neq m$). Besides, the integral ${\mathscr{I}}^{nm}_2$ is nonvanishing only for $n=m$ and becomes unity for $r_y = r_f$. At this point we would like to mention that, in our analysis in order to evade the complicacy of mode mixing and construct a simpler form of fermion mixing matrix we choose the condition of equality  ($r_y$=$r_f$)  \cite{Jha:2014faa, Datta:2015aka, Datta:2016flx, Shaw:2019fin, Biswas:2017vhc}. This equality condition\footnote{However, in general, one can proceed with unequal coefficients of boundary terms for kinetic and Yukawa interaction for fermions.} ($r_y=r_f$) has been maintained in the rest of our analysis. 

With the above mentioned equality condition ($r_y=r_f$), the mass matrix (given in Eq.\;\ref{fermion_mix}) can easily be diagonalised by the following bi-unitary transformations for the left- and right-handed fields \cite{Datta:2015aka, Datta:2016flx, Shaw:2019fin, Biswas:2017vhc}
\begin{equation}
U_{L}^{(n)}=\begin{pmatrix}
\cos\alpha_{tn} & \sin\alpha_{tn} \\ -\sin\alpha_{tn} & \cos\alpha_{tn}
\end{pmatrix},~~U_{R}^{(n)}=\begin{pmatrix}
\cos\alpha_{tn} & \sin\alpha_{tn} \\ \sin\alpha_{tn} & -\cos\alpha_{tn}
\end{pmatrix},
\end{equation}
where, $\alpha_{tn}\left[ = \frac12\tan^{-1}\left(\frac{m_{t}}{m_{f^{(n)}}}\right)\right]$ is identified as the mixing angle. The gauge eigen states $\Psi_L(x,y)$ and $\Psi_R(x,y)$ can be expressed in terms of mass eigen states $T^1_t$ and $T^2_t$ by the following relations \cite{Datta:2015aka, Datta:2016flx, Shaw:2019fin, Biswas:2017vhc}

\vspace*{-1cm}
\begin{tabular}{p{8cm}p{8cm}}
{\begin{align}
&{\phi^{(n)}_L} =  \cos\alpha_{tn}T^{1(n)}_{tL}-\sin\alpha_{tn}T^{2(n)}_{tL},\nonumber \\
&{\chi^{(n)}_L} =  \cos\alpha_{tn}T^{1(n)}_{tR}+\sin\alpha_{tn}T^{2(n)}_{tR},\nonumber
\end{align}}
&
{\begin{align}
&{\phi^{(n)}_R} =  \sin\alpha_{tn}T^{1(n)}_{tL}+\cos\alpha_{tn}T^{2(n)}_{tL},\nonumber \\
&{\chi^{(n)}_R} =  \sin\alpha_{tn}T^{1(n)}_{tR}-\cos\alpha_{tn}T^{2(n)}_{tR}.
\end{align}}
\end{tabular}
Both the mass eigen states $T^{1(n)}_t$ and $T^{2(n)}_t$ have the identical mass eigen value at each KK-level. For $n^{th}$ KK-level the mass eigen value takes the form as $M_{t^{(n)}} \equiv \sqrt{m_{t}^{2}+m^2_{f^{(n)}}}$.

Let us look at the kinetic actions (governed by $SU(2)_L \times U(1)_Y$ gauge group) of 5D gauge and scalar fields including their corresponding BLKTs \cite{Flacke:2008ne, Datta:2014sha, Jha:2014faa, Datta:2015aka, Datta:2016flx, Shaw:2019fin, Biswas:2017vhc, Dey:2016cve}
\begin{eqnarray}
S_{gauge} &=& -\frac{1}{4}\int d^5x \bigg[ W^a_{MN} W^{aMN}+r_W \left\{ \delta(y) +  \delta(y - \pi R)\right\} W^a_{\mu\nu} W^{a\mu \nu}\nonumber \\
&+& B_{MN} B^{MN}+r_B \left\{ \delta(y) +  \delta(y - \pi R)\right\} B_{\mu\nu} B^{\mu \nu}\bigg],
\label{pure-gauge}
\end{eqnarray}
\vspace*{-1cm}
\begin{eqnarray} 
S_{scalar} &=& \int d^5x \bigg[ (D_{M}\Phi)^\dagger(D^{M}\Phi) + r_\phi \left\{ \delta(y) +  \delta(y - \pi R)\right\} (D_{\mu}\Phi)^\dagger(D^{\mu}\Phi) \bigg],
\label{higgs}
\end{eqnarray}
where, $r_W$, $r_B$ and $r_\phi$ are designated as the coefficients of the BLKTs for the respective fields. 5D field strength tensors are expressed as
\begin{eqnarray}\label{ugfs}
W_{MN}^a &\equiv& (\partial_M W_N^a - \partial_N W_M^a-{\tilde{g}_2}\epsilon^{abc}W_M^bW_N^c),\\ \nonumber
B_{MN}&\equiv& (\partial_M B_N - \partial_N B_M).
\end{eqnarray}
$W^a_M (\equiv W^a_\mu, W^a_4)$ and $B_M (\equiv B_\mu, B_4)$ ($M=0,1 \ldots 4$) are considered as the 5D gauge fields corresponding to the gauge groups $SU(2)_L$ and $U(1)_Y$ respectively. 5D covariant derivative can be written as $D_M\equiv\partial_M+i{\tilde{g}_2}\frac{\sigma^{a}}{2}W_M^{a}+i{\tilde{g}_1}\frac{Y}{2}B_M$, where, ${\tilde{g}_2}$ and ${\tilde{g}_1}$ are represented as the 5D gauge coupling constants. Generators of $SU(2)_L$ and $U(1)_Y$ gauge groups are represented by $\frac {\sigma^{a}}{2} (a\equiv 1\ldots 3)$ and $\frac Y2$ respectively. Each of the gauge and scalar fields which are involved in the above actions (Eqs.\;\ref{pure-gauge} and \ref{higgs}) can be manifested using appropriate KK-wave-functions as \cite{Datta:2014sha, Jha:2014faa, Datta:2015aka, Datta:2016flx, Shaw:2019fin, Biswas:2017vhc, Dey:2016cve}
\begin{equation}\label{Amu}
V_{\mu}(x,y)=\sum_n V_{\mu}^{(n)}(x) a^n(y),\;\;\;\;\
V_{4}(x,y)=\sum_n V_{4}^{(n)}(x) b^n(y)
\end{equation}
\vspace*{-0.5cm}
and
\begin{equation}\label{chi}
\Phi(x,y)=\sum_n \Phi^{(n)}(x) h^n(y),
\end{equation}
where both the 5D $SU(2)_L$ and $U(1)_Y$ gauge bosons are generically illustrated by $(V_\mu, V_4)$.
 
To this end, we would like to discuss some important points by which one can understand the following gauge and scalar field structure as well as the corresponding KK-wave-functions. The physical neutral gauge bosons emerge from the mixing of $B$ and $W^3$ fields and hence the KK-decompositions of neutral gauge bosons are very complicated in the current extra dimensional scenario due to the presence of two types of mixings both at the bulk as well as on the boundary. Therefore, under this circumstances, it would be very difficult to diagonalise the bulk and boundary actions simultaneously by the same 5D field redefinition\footnote{However, in general one can proceed with $r_W\neq r_B$, but in this case the mixing between $B$ and $W^3$ in the bulk and on the boundary points generate off-diagonal terms in the neutral gauge boson mass matrix.} except the condition $r_W=r_B$. Hence, in the following, we will keep the equality condition $r_W=r_B$ \cite{Datta:2014sha, Jha:2014faa, Datta:2015aka, Datta:2016flx, Shaw:2019fin, Biswas:2017vhc, Dey:2016cve} and as a consequence we obtain the same structure (like mUED scenario) of mixing between KK-excitations of the neutral component of the gauge fields (i.e., the mixing between  $B^{(n)}$ and $W^{3(n)}$) in nmUED scenario. Thereafter, the mixing between $B^{(1)}$ and $W^{3(1)}$ (i.e., the mixing at the first KK-level) provides the $\gamma^{(1)}$ and $Z^{(1)}$. This $\gamma^{(1)}$ (first excited KK-state of the photon) is completely stable due to the conservation of KK-parity and it acquires the lowest mass among the first excited KK-states in the nmUED particle spectrum. Furthermore, it could not decay to pair of SM particles. Therefore, this $\gamma^{(1)}$ can be considered as a viable DM candidate in this scenario \cite{Datta:2013nua}.  

Let us discuss on the gauge fixing action (considering a generic BLKT parameter $r_V$ for gauge bosons) for nmUED scenario \cite{Datta:2014sha, Jha:2014faa, Datta:2015aka, Datta:2016flx, Shaw:2019fin, Biswas:2017vhc, Dey:2016cve} 
\begin{eqnarray} 
S_{gauge\;fixing} &=& -\frac{1}{\xi _y}\int d^5x\Big\vert\partial_{\mu}W^{\mu +}+\xi_{y}(\partial_{y}W^{4+}+iM_{W}\phi^{+}\{1 + r_{V}\left( \delta(y) + \delta(y - \pi R)\right)\})\Big \vert ^2 \nonumber \\&&-\frac{1}{2\xi_y}\int d^5x [\partial_{\mu}Z^{\mu}+\xi_y(\partial_{y}Z^{4}-M_{Z}\chi\{1+ r_V( \delta(y) +  \delta(y - \pi R))\})]^2\nonumber \\
&-&\frac{1}{2\xi_y}\int d^5x [\partial_{\mu}A^{\mu}+\xi_y\partial_{y}A^{4}]^2, 
\label{gauge-fix}
\end{eqnarray}
where $M_W(M_Z)$ is identified as the mass of the SM $W^\pm (Z)$ boson.
For an extensive study on gauge fixing action/mechanism in nmUED we refer to \cite{Datta:2014sha}. The above action (given in Eq.\;\ref{gauge-fix}) is very intricate and at the same time very important for this nmUED scenario where we will compute one-loop diagrams (necessary for present calculation) in Feynman gauge. Due to the impact of the BLKTs, the Lagrangian leads to a non-homogeneous weight function for the fields with respect to the extra dimension. This inhomogeneity enforces us to define a $y$-dependent gauge fixing parameter $\xi_y$ as \cite{Datta:2014sha, Jha:2014faa, Datta:2015aka, Datta:2016flx, Shaw:2019fin, Biswas:2017vhc, Dey:2016cve}
\begin{equation}\label{gf_para}
\xi =\xi_y\,(1+ r_V\{ \delta(y) +  \delta(y - \pi R)\}),
\end{equation}
where $\xi$ is independent of $y$. The above relation behaves as {\em renormalisation} of the gauge fixing parameter since the BLKTs are in some sense contributed as the counterterms taking into account the unknown ultraviolet correction in loop calculations. With this  in mind, we treat $\xi_y$ as the bare gauge fixing parameter while $\xi$ can be viewed as the renormalised gauge fixing parameter taking the values $0$ (Landau gauge), $1$ (Feynman gauge) or $\infty$ (Unitary gauge) \cite{Datta:2014sha}.

In this nmUED scenario appropriate gauge fixing procedure enforces the equality condition of coefficient of the BLKTs of gauge and scalar fields, i.e., $r_V=r_\phi$~\cite{Datta:2014sha, Jha:2014faa, Datta:2015aka, Datta:2016flx, Shaw:2019fin, Biswas:2017vhc, Dey:2016cve}. In this limit, KK-masses for the gauge and the scalar fields are equal ($m_{V^{(n)}}(=m_{\phi^{(n)}})$) and can be derived from the same transcendental equation (Eq.~\ref{fermion_mass}). At the $n^{th}$ KK-level the physical gauge fields ($W^{\mu (n)\pm}$) and charged Higgs ($H^{(n)\pm}$) have the same\footnote{In the same way one can find the mass eigen values for the KK-excited $Z$ boson and pseudo scalar $A$. Also, their mass eigen values are identical to each other at any KK-level. For example at $n^{th}$ KK-level it takes the form as $\sqrt{M_{Z}^{2}+m^2_{V^{(n)}}}$. } mass eigen value and is given by\cite{Datta:2014sha, Jha:2014faa, Datta:2015aka, Datta:2016flx, Shaw:2019fin, Biswas:2017vhc, Dey:2016cve} 
\be\label{MWn}
M_{W^{(n)}} = \sqrt{M_{W}^{2}+m^2_{V^{(n)}}}\;.
\ee 
Besides, in the ’t-Hooft Feynman gauge, the mass of Goldstone bosons ($G^{(n)\pm}$) corresponding to the gauge fields $W^{\mu (n)\pm}$ have the same value $M_{W^{(n)}}$\cite{Datta:2014sha, Jha:2014faa, Datta:2015aka, Datta:2016flx, Shaw:2019fin, Biswas:2017vhc, Dey:2016cve}.
 
To this end, we would like to focus on the interactions that will involve in our calculation. We can derive these interaction by integrating out the 5D action over the extra space-like dimension ($y$) using the specific $y$-dependent KK-wave-function for the respective fields in 5D action. Consequently,  some of the interactions are modified by so called overlap integrals with respect to their mUED counterparts. The actual form of the overlap integrals have been given in Appendix \ref{fyerul}. Detailed discussions on these overlap integrals have been given in \cite{Datta:2015aka}. 
\section{\boldmath{$\Delta B=2$} transitions in nmUED scenario}\label{bbbar_mixing}
The effective Hamiltonian which governs the $\Delta B=2$ transitions in the SM\cite{Buras:1990fn, Urban:1997gw} can easily be modified for the nmUED scenario as follows
\begin{eqnarray}\label{hdb2}
{\cal H}^{\Delta B=2}_{\rm eff}&=&\frac{G^2_{\rm F}}{16\pi^2}M^2_W
 \left(V^\ast_{tb}V_{tq}\right)^2 \eta_{B}
 S(x_t,r_f,r_V,R^{-1})\left[\alpha^{(5)}_s(\mu_b)\right]^{-6/23}\left[
  1 + \frac{\alpha^{(5)}_s(\mu_b)}{4\pi} J_5\right] \\ \nonumber
&&[\bar b\gamma_\mu (1-\gamma_5)q][\bar b\gamma^\mu (1-\gamma_5)q] + {\rm h. c.},
\end{eqnarray}
with $q=d,s$. Here $\mu_b=\ord(m_b)$, $J_5=1.627$ and 
\begin{equation}\label{etb}
\eta_B=0.55\pm0.01 
\end{equation}
represents the short distance Quantum Chromo Dynamics (QCD) corrections \cite{Buras:1990fn,Urban:1997gw}. 
The function $S(x_t,r_f,r_V,R^{-1})$ represents the total contribution in nmUED scenario as given below
\be\label{SACD}
S(x_t,r_f,r_V,R^{-1})=S_0(x_t)+\sum_{n=1}^\infty S_n(x_t,x_{f^{(n)}},x_{V^{(n)}})\;,
\ee
with $x_t=\frac{m^2_t}{M^2_W}$, $x_{V^{(n)}}=\frac{m^2_{V^{(n)}}}{M^2_W}$ and $x_{f^{(n)}}=\frac{m^2_{f^{(n)}}}{M^2_W}$. $m_{V^{(n)}}$ and $m_{f^{(n)}}$ can be obtained from transcendental equation given in Eq.\;\ref{fermion_mass}. Moreover, 
\begin{equation}\label{S0}
S_0(x_t)=\frac{4x_t-11x^2_t+x^3_t}{4(1-x_t)^2}-
 \frac{3x^3_t \ln x_t}{2(1-x_t)^3}
\end{equation}
depicts the SM contribution \cite{Buras:2002ej, Buras:1994ec, Buras:1998raa} and is obtained from the box diagrams with $(W^\pm,t)$ and $(G^\pm,t)$ exchanges with the $m_t$ independent terms  eliminated by the GIM mechanism.

\begin{figure}[htbp!]
\begin{center}
\subfloat[]{\includegraphics[height=4cm,width=9cm,angle=0]{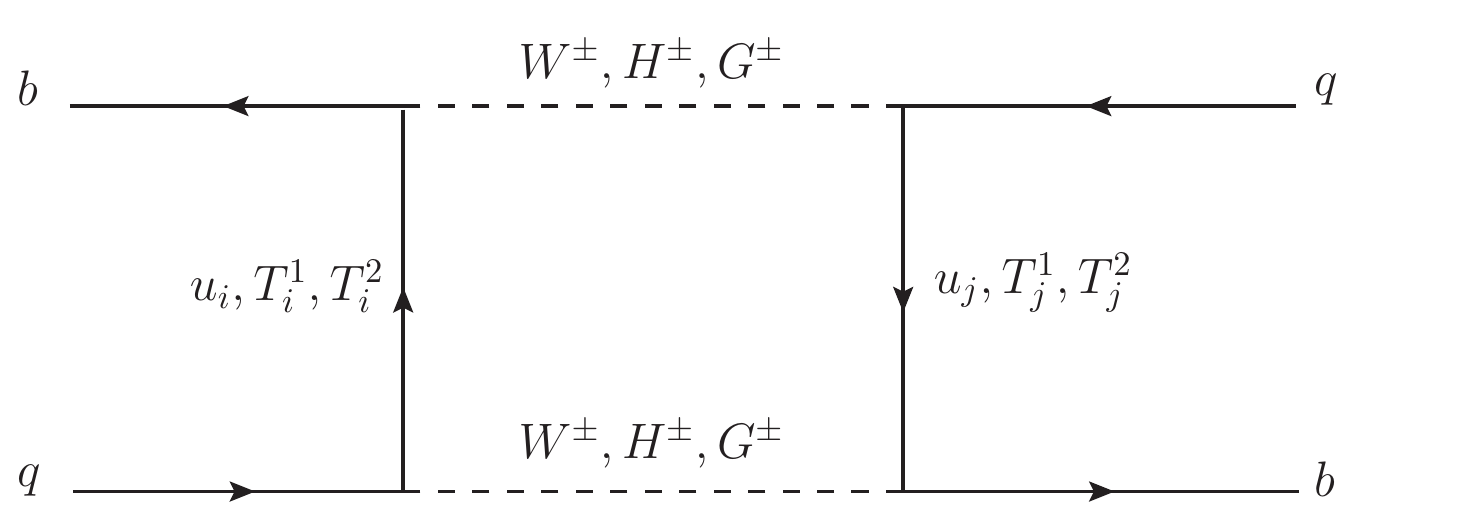}}
\subfloat[]{\includegraphics[height=4cm,width=9cm,angle=0]{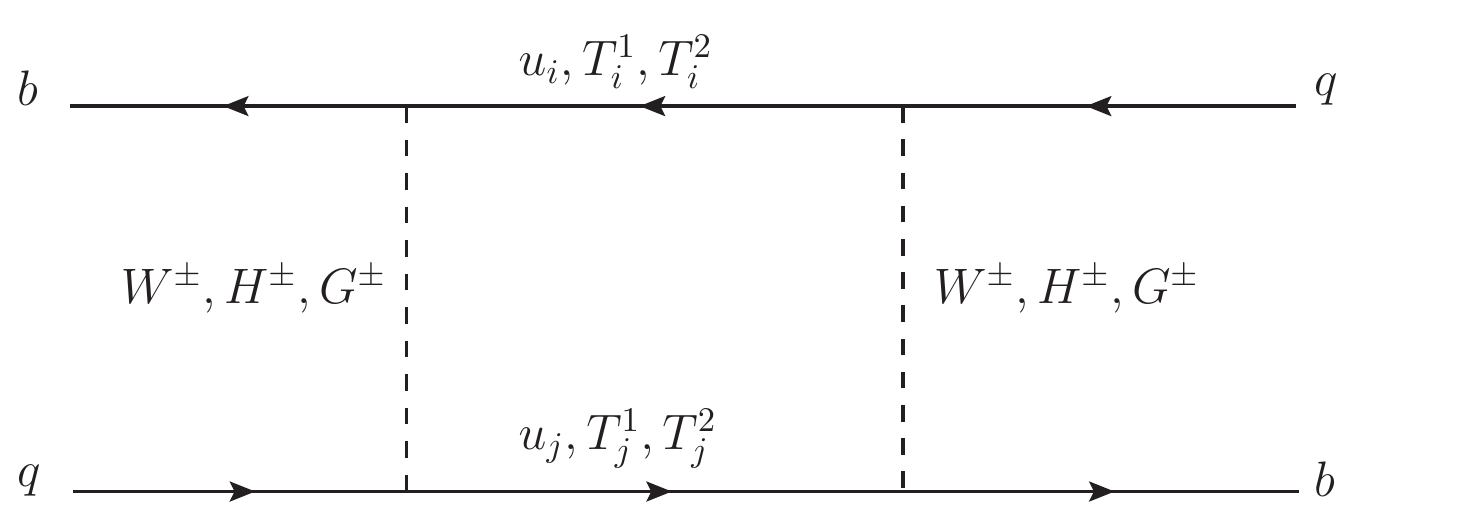}}
\caption{The relevant box diagrams contributing to $S_n(x_t,x_{f^{(n)}},x_{V^{(n)}})$ in nmUED scenario.} 
\label{bb_box}
\end{center}
\end{figure}

The function $S_n(x_t,x_{f^{(n)}},x_{V^{(n)}})$ indicate the KK-contributions which is obtained by evaluating the box diagrams shown in Fig.~\ref{bb_box} with 
$W^{(n)\pm}$, $H^{(n)\pm}$, $G^{(n)\pm}$, 
$T^{1(n)}_i$ and $T^{2(n)}_i$ ($i=u,c,t$)
interchanges and  multiplying the result by $i/4$, where $1/4$ is a 
combinatorial factor. Momenta and masses of external quarks have been neglected in this calculation. Using the unitarity condition  of CKM matrix we can write the functions $S_n(x_t,x_{f^{(n)}},x_{V^{(n)}})$ as \cite{Buras:2002ej}
\be\label{SN}
S_n(x_t,x_{f^{(n)}},x_{V^{(n)}})\equiv F(x_{t^{(n)}},x_{t^{(n)}})+F(x_{u^{(n)}},x_{u^{(n)}})
-2 F(x_{t^{(n)}},x_{u^{(n)}})~,
\ee
where the function $F(x_{i^{(n)}},x_{j^{(n)}})$ is representing the sum of the contribution of the 
diagrams corresponding to a given pair $(m_{i(n)},m_{j(n)})$ to $S_n(x_t,x_{f^{(n)}},x_{V^{(n)}})$, where 
\be\label{xin}
x_{i(n)}=\frac{m^2_{i(n)}}{M_{W^{(n)}}^2}=\frac{m^2_i+ m^2_{f^{(n)}}}{M^2_W+m^2_{V^{(n)}}}\;.
\ee
Finally the compact form of this function obtained from one-loop box diagrams (see Fig.\;\ref{bb_box}) is given by 
\begin{eqnarray}\label{SFIN}
S_n(x_t,x_{f^{(n)}},x_{V^{(n)}})&=&\frac14\Bigg[\frac{1}{(-1+x_{f^{(n)}}-x_{V^{(n)}})^3}\bigg[(-1+x_{f^{(n)}}-x_{V^{(n)}})\bigg\{(I^n_2)^4(1+x_{f^{(n)}}\\ \nonumber
&&-15x_{V^{(n)}}) +4(I^n_1)^4(1+x_{f^{(n)}}+x_{V^{(n)}})
\bigg\}-2\bigg\{4(I^n_1)^4x_{f^{(n)}}(1+x_{V^{(n)}})\\ \nonumber
&&+(I^n_2)^4\left(x_{f^{(n)}}-3x_{f^{(n)}}x_{V^{(n)}}-4x_{V^{(n)}}(1+x_{V^{(n)}})\right)\bigg\}\ln\bigg(\frac{x_{f^{(n)}}}{1+x_{V^{(n)}}}\bigg) 
\bigg]\\ \nonumber
&&-2\bigg[ \frac{\bigg\{(I^n_2)^4(1-7x_{V^{(n)}})+4(I^n_1)^4(1+x_{V^{(n)}})\bigg\}}{(-1+x_{f^{(n)}}-x_{V^{(n)}})(-1+x_t+x_{f^{(n)}}-x_{V^{(n)}})}\\ \nonumber
&&-\frac{x_{f^{(n)}}\bigg\{4(I^n_1)^4x_{f^{(n)}}+(I^n_2)^4(x_{f^{(n)}}-8x_{V^{(n)}})\bigg\}}{x_t(-1+x_{f^{(n)}}-x_{V^{(n)}})^2}\ln\bigg(\frac{x_{f^{(n)}}}{1+x_{V^{(n)}}}\bigg)\\ \nonumber
&&+\frac{\bigg(1+\frac{x_{f^{(n)}}}{x_t}\bigg)\bigg\{4(I^n_1)^4(x_t+x_{f^{(n)}})+(I^n_2)^4(x_t+x_{f^{(n)}}-8x_{V^{(n)}})\bigg\}}{(-1+x_t+x_{f^{(n)}}-x_{V^{(n)}})^2}\times\\ \nonumber
&&\ln\bigg(\frac{x_{f^{(n)}}+x_t}{1+x_{V^{(n)}}}\bigg)
\bigg]+\frac{1}{(-1+x_t+x_{f^{(n)}}-x_{V^{(n)}})^3}\bigg[(-1+x_t+x_{f^{(n)}}-x_{V^{(n)}})\\ \nonumber
&&\bigg\{(I^n_2)^4(1+x_t+x_{f^{(n)}}-15x_{V^{(n)}})+(I^n_1)^4\bigg(x_{f^{(n)}}(4+x^2_t)\\ \nonumber
&&+4(1+x_{V^{(n)}})+x_t\left(4+x_t(-15+x_t+x_{V^{(n)}})\right)\bigg)\bigg\}\\ \nonumber
&&-2\bigg\{(I^n_2)^4\left(x_{f^{(n)}}+x_t-4x_{V^{(n)}}-3(x_{f^{(n)}}+x_t)x_{V^{(n)}}-4x^2_{V^{(n)}}\right)\\ \nonumber
&&+(I^n_1)^4\bigg(x_t\left(4-4x_t-3x^2_t+(x_t-2)^2x_{V^{(n)}}\right) \\ \nonumber
&&+x_{f^{(n)}}\left(x^2_t(-3+x_{V^{(n)}})+4(1+x_{V^{(n)}})\right)
\bigg)
\bigg\}\ln\bigg(\frac{x_{f^{(n)}}+x_t}{1+x_{V^{(n)}}}\bigg)
\bigg]\Bigg].
\end{eqnarray}
With the increasing value of KK-modes (i.e., with higher value $n$) the masses of the fields $T^{1(n)}_t$, $T^{2(n)}_t$, $T^{1(n)}_u$ and $T^{2(n)}_u$ become degenerate in nature
and consequently the function $S_n(x_t,x_{f^{(n)}},x_{V^{(n)}})$ diminishes with larger values of $n$. Therefore, only a few terms in the sum given in Eq.\;\ref{SACD} are relevant.  Here, $I^n_1$ and $I^n_2$ are two overlap integrals and their expressions are given in the Appendix\;\ref{fyerul}. 

{It has already been addressed that, due to the presence of different BLTs in the nmUED
action, the KK-masses and couplings (modified by $I^n_1$ and $I^n_2$) involving KK-excitations
are nontrivially modified with respect to their UED counterparts. Therefore, it would not be possible to evaluate the
function $S_n$ in nmUED scenario simply by
rescaling the same of the UED model \cite{Buras:2002ej}. Hence, we have computed the function $S_n$ independently using the box diagram (Fig.\;\ref{bb_box}) for the nmUED scenario.
Moreover, it is quite clear from Eq.\;\ref{SFIN} that the function $S_n$ is drastically different from that of the UED
expression. However, if we set the boundary
terms to zero; i.e., $r_V=0$ and $r_f=0$, then we can easily reproduce the result of the UED version from our expression. Further, we would like to mention that in our
computation of one-loop box diagrams we consider only those interactions in which
zero-mode field couples to a pair of KK-excitations with
equal KK-numbers. Moreover, in the KK-parity conserving nmUED scenario one can also have nonzero 
interactions containing KK-excitations with KK-numbers $n$, $m$ and
$p$, where $n+m+p$ is an even integer. However, we have
explicitly verified that the final results would not change
significantly even if one considers the contributions of all the
possible off-diagonal interactions \cite{Jha:2014faa, Datta:2015aka, Datta:2016flx, Shaw:2019fin}.

To this end, applying the same procedure as in the SM, we can calculate the mass differences $\Delta M_q$ for nmUED scenario. For this purpose, we can readily adopt the
technique of UED scenario as given in \cite{Buras:2002ej}, because the basic structures (also the structure of operator responsible for $B_q$-mixing) of both UED and nmUED scenarios are similar. Therefore, all other physical aspects (apart from the function $S_n$ obtained from the one-loop box diagrams) of both the UED and nmUED scenarios are same. Hence, one can easily write the expression of mass difference in nmUED scenario as}
\begin{equation}\label{DMQ}
\Delta M_q = \frac{G_{\rm F}^2}{6 \pi^2} \eta_B m_{B_q} 
(\hat B_{B_q} F_{B_q}^2 ) M_W^2 S(x_t,r_f,r_V,R^{-1}) |V_{tq}|^2\;.
\end{equation}
Here, $F_{B_q}$ represents the $B_q$-meson decay constant and
$\hat B_q$ the renormalisation group invariant parameter related 
to the hadronic matrix element of the operator $\Delta B=2$ \cite{Buras:2001pn}.

At this moment, we would like to make a few comments on the QCD factor $\eta_B$ 
(given in Eq.\;\ref{etb}) which has been evaluated within the SM 
including next-to-leading oder (NLO) QCD corrections. These are necessary for the proper matching
of the Wilson Coefficient (WC) of the operator ($\Delta B=2$) with its hadronic 
matrix element designated by the parameter $\hat B_{B_{s,d}}$ and evaluated using non-perturbative methods. 
Since the KK-modes and the top quark are 
integrated out at a single scale $\mu_t=\ord(\mt,R^{-1})$, therefore at a scale lower than $\mu_t$, the contributions 
to $\eta_B$ for the nmUED scenario and for the SM are same. They just represent the finite renormalisation of the operator $(\Delta B=2)$
from the scales $\ord(\mu_t)$ down to the scales $\ord(m_b)$. The disagreement in  
QCD corrections between the KK-contributions and the SM contributions appears only in the full theory at scales $\mu_t=\ord(\mt,R^{-1})$. In this situation the unknown QCD corrections to the box diagrams in Fig. \ref{bb_box} can, in principle,
differ from the known QCD corrections to the SM box diagrams \cite{Buras:1990fn, Urban:1997gw} that have 
been included in $\eta_B$. However, as the QCD coupling constant $\alpha_s(\mu_t)$ is small and the 
QCD corrections to the SM box diagrams are of order of a few percent, therefore one can expect that the difference between the QCD corrections to the diagrams in Fig.~\ref{bb_box}  and to the SM box diagrams
is insignificant \cite{Buras:2002ej}. Therefore, in the following, we will use the same QCD factor and hadronic matrix element in SM as well as in the concerned nmUED scenario. With this we will study the $\Delta B=2$ transitions in nmUED scenario and try to estimate the impact of BLT parameters on the CKM elements and UT.

{\subsection{Effects of BLTs on CKM parameters and Unitarity Triangle in the nmUED scenario}\label{ckmut}
In nmUED scenario, we evaluate the KK-contributions to the function $S_n$ which is significantly affected by BLT parameters $(r_V, r_f)$. Therefore, using this function $S_n$ we can study the effects of BLT parameters on the elements of the CKM matrix and in particular on the shape of the UT. However, in order to execute this strategy we need to recall some features of CKM matrix and the UT as depicted in Fig.~\ref{fig:utriangle}. Using the Wolfenstein parametrisation \cite{Wolfenstein:1983yz} as generalised to higher orders\footnote{One should note that here, terms greater than $\mathcal{O}(\lambda^4)$ have been neglected. However, this may give wrong results. In order to improve the accuracy of the UT a correction term $\approx \mathcal{O}(\lambda^5)$  has been included to $V_{td}$ element.} in $\lambda$, the CKM matrix can be written as \cite{Buras:1994ec}
\begin{eqnarray}
{\rm V}_{\rm CKM} = \left(\begin{array}{ccc}
1-\frac{\lambda^2}{2} & \lambda & \mathbb{A}\lambda^3(\rho-i\eta) \\
-\lambda & 1-\frac{\lambda^2}{2} & \mathbb{A}\lambda^2 \\
\mathbb{A}\lambda^3[1-(\rho+i\eta)(1-\frac{\lambda^2}{2})]  & -\mathbb{A}\lambda^2 &  1 \end{array}
\right)\,. \nonumber \\ 
\label{VCKM}
\end{eqnarray}
Here $\lambda$, $\mathbb{A}$, $\rho$ and $\eta$ are the Wolfenstein 
parameters \cite{Wolfenstein:1983yz}. The most commonly used UT arises from
\begin{equation}
V_{ud}V^*_{ub}+V_{cd}V^*_{cb}+V_{td}V^*_{tb}=0\;,
\end{equation}
by dividing each side by the best known one, $V_{cd}V^*_{cb}$. Its vertices are
exactly $C=(0, 0)$, $B=(1, 0)$, and $A=(\bar{\rho}, \bar{\eta})$, where $\bar{\rho}$ and  
$\bar{\eta}$ of the UT is given by \cite{Buras:1994ec}
\begin{equation}\label{2.88d}
\bar\rho=\rho (1-\frac{\lambda^2}{2}),
\qquad
\bar\eta=\eta (1-\frac{\lambda^2}{2}).
\end{equation}

\begin{figure}[htbp!]
\begin{center}
\includegraphics[height=6cm,width=11cm,angle=0]{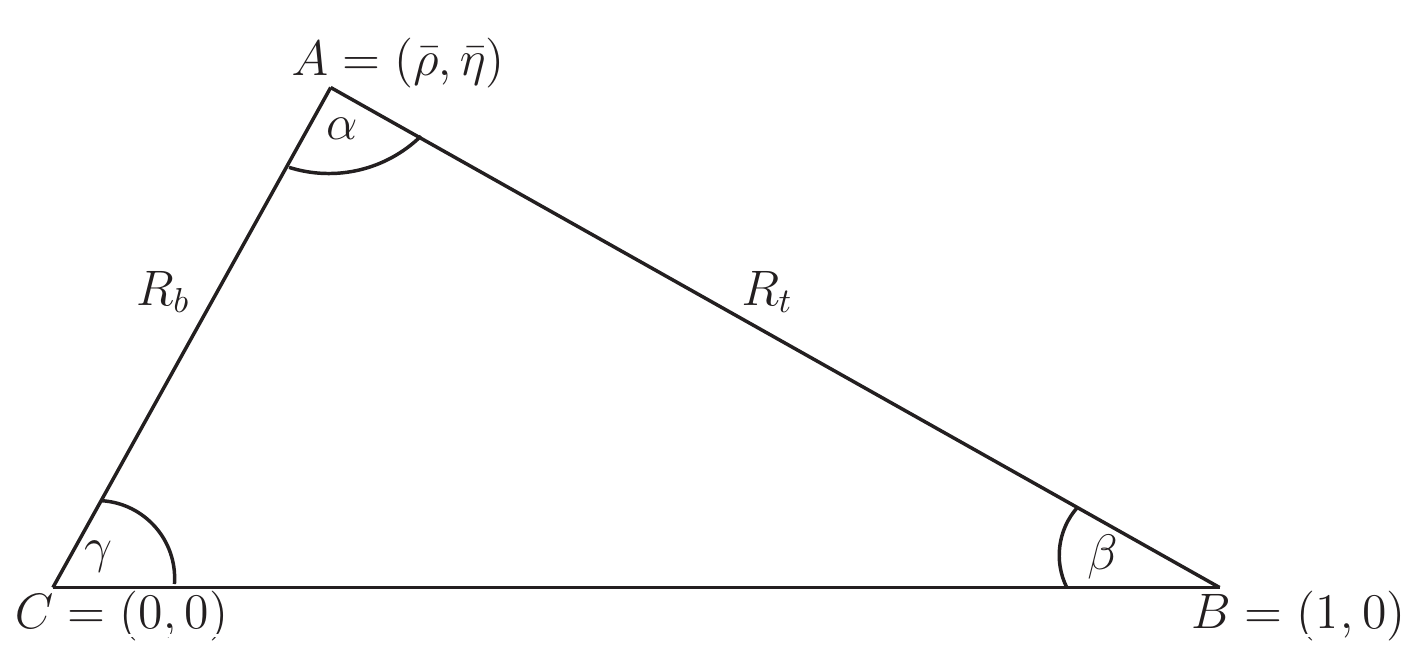}
\caption{Unitarity Triangle} 
\label{fig:utriangle}
\end{center}
\end{figure}
The lengths $R_b$ and $R_t$ can be expressed as \cite{Buras:2002ej, Buras:1994ec}
\begin{equation}\label{2.94}
R_b \equiv \frac{| V_{ud}^{}V^*_{ub}|}{| V_{cd}^{}V^*_{cb}|}
= \sqrt{\bar\rho^2 +\bar\eta^2}
= (1-\frac{\lambda^2}{2})\frac{1}{\lambda}
\left| \frac{V_{ub}}{V_{cb}} \right|\;,
\end{equation}
\begin{equation}\label{2.95}
R_t \equiv \frac{| V_{td}^{}V^*_{tb}|}{| V_{cd}^{}V^*_{cb}|} =
 \sqrt{(1-\bar\rho)^2 +\bar\eta^2}
=\frac{1}{\lambda} \left| \frac{V_{td}}{V_{cb}} \right|\;,
\end{equation}
while the angles $\gamma$ and  $\beta$ of the UT are connected
directly to the complex phases of the CKM elements $V_{td}$ and
$V_{ub}$, respectively, via
\be\label{e417}
\quad V_{ub}=|V_{ub}|e^{-i\gamma}\;, V_{td}=|V_{td}|e^{-i\beta}\;.
\ee

The value of $R_b$ (given in \ref{2.94}), i.e., the length of the side AC 
is determined from $\vub$. $\vcb$ and $|V_{ub}|$ are in general determined from tree level decays. Moreover, in the nmUED scenario there are no KK-contributions at the tree level, hence
absolute values of these CKM elements of nmUED scenario are approximately same as that of the SM. Besides, if we look from the perspective of the UT, the lengths of its two sides, 
AC and CB are common to the SM and the nmUED scenario. Moreover, we would like to mention that in the nmUED scenario as there 
are no new complex phases beyond the KM phase, the angle $\beta$ as extracted by means of $a_{\psi K_S}$ in $B\to \psi K_s$ is common to both nmUED scenario and SM. The world average value of the angle $\beta$ which we will use in our following analysis is given below \cite{Amhis:2019ckw}
\be
\beta=(22.2\pm 0.7)^\circ~.
\label{ga}
\ee

Before proceeding further, we would like to mention that, in view of the previous discussions we can easily categorise the present nmUED scenario as an MFV model \cite{Buras:2002yj,DAmbrosio:2002vsn}. The reason is that this nmUED scenario (like UED scenario \cite{Buras:2002ej}) is a class of extensions of the SM in which only the SM operators
in the effective weak Hamiltonian are relevant and flavour violation is entirely governed by the CKM matrix. Moreover, CP violation is governed solely by the KM phase. Besides, one of the fascinating features of the MFV models is the existence of the universal UT (UUT) \cite{Buras:2000dm} that can be constructed from quantities in which all the
dependence on NP cancels out or is negligible like in tree level decays. Now, in spite of the existence of  common UUT for both the nmUED and SM scenarios, a crucial difference is $S(x_t,r_f,r_V,R^{-1})\not=S_0(x_t)$. Therefore, only one from the following observables: $\varepsilon_K$\footnote{In view of the discussions given in ref. \cite{Buras:2002ej} the effect of the KK-modes on the charm- and mixed charm-top contributions is totally insignificant. Considerable effect for these modes are only top contributions which has been indicated by the same function $S(x_t,r_f,r_V,R^{-1})$ as in the case of $\Delta M_q$. Therefore, the $K_L-K_S$ mass difference $\Delta M_K$, is practically dominated by internal charm contributions and essentially unaffected by the KK-modes. Consequently, in the present nmUED scenario we have also found the same phenomenon. We hence refrain from providing the details of $\varepsilon_K$ in the present article.}, $\Delta M_d$ and $\Delta M_s$ will agree with the experimental data for only one of the two scenarios (SM and nmUED). 

At this stage, considering the above mentioned facts, we can present a picture which shows how the quantities $\vtd$, $\bar\rho$, $\bar\eta$ and $\gamma$ are affected by the BLT parameters ($r_V$, $r_f$) and $R^{-1}$. In order to obtain these results in nmUED scenario we adopt the following procedure which is quite similar as given in \cite{Buras:2002ej} {\it albeit} for UED scenario. At first we use the $B^0_d-\bar B^0_d$ mixing constraint $\Delta M_d$. Using the central value of experimental data of $\Delta M_d=(0.5065\pm 0.0019){\rm ps}^{-1}$\cite{Zyla:2020zbs, Amhis:2019ckw} and utilising the function given in Eq.\;\ref{SACD} we can determine\footnote{Here, we use $\sqrt{\hat B_{B_d}}F_{B_d}=(225\pm 9)$ MeV \cite{Zyla:2020zbs} and $m_{B_d}=(5279.63\pm 0.20)$ MeV \cite{Zyla:2020zbs}.} the value of $\vtd$ in nmUED scenario from Eq.\;\ref{DMQ}. After that, using this $\vtd$ and with the allowed value of $\lambda=0.224837^{+0.000251}_{-0.000060}$ \cite{CKM:summer} and $V_{cb}=(41.0\pm 1.4)\times 10^{-3}$ \cite{Zyla:2020zbs}
we will find $R_{t}$ via Eq.\;\ref{2.95}. Finally, using the following relations \cite{Buras:2002yj}
\be
\bar{\eta}=R_t\sin(\beta),\quad \bar{\rho}=1-R_t\cos(\beta) \quad {\rm and} \quad \cot{\gamma}=\frac{1-R_t\cos(\beta)}{R_t\sin(\beta)},
\label{rhoetagama}
\ee
we will compute the values of  
$\bar\rho$, $\bar\eta$ and $\gamma$ respectively, as functions of $R^{-1}$ and BLT parameters ($r_V$, $r_f$) in nmUED scenario. With this procedure we can obtain the parameter space which satisfies 2$\sigma$ allowed range of the values of the quantities ($\vtd=(8.0\pm 0.6)\times 10^{-3}$\cite{Zyla:2020zbs}, $\bar\rho=0.157^{+0.027}_{-0.012}$ \cite{CKM:summer}, $\bar\eta=0.350^{+0.018}_{-0.016}$ \cite{CKM:summer} and $\gamma=(72.1\pm 8.6)^\circ$\cite{Zyla:2020zbs}). As, for $\bar{\rho}$ and $\bar{\eta}$, the distributions are perfectly Gaussian \cite{CKM:summer}, therefore, we can definitely take the values with 2$\sigma$ uncertainties for $\bar{\rho}$ and $\bar{\eta}$ with all other observables for which 2$\sigma$ uncertainties are given as twice the 1$\sigma$ uncertainty. 

Additionally, with the above mentioned conditions we would also like to impose another condition simultaneously by studying the effects of BLT parameters ($r_V$, $r_f$) and $R^{-1}$ on $\Delta M_s$. For this purpose, we would like to mention that $\vts$ is very close to $\vcb$ due to CKM unitarity. As $\vcb$ is common for both the SM and nmUED scenario, therefore, $\vts$ is common with an excellent accuracy to both models and consequently we can have the following relation\footnote{In view of the discussions given just before the subsection \ref{ckmut}, we assume that all other quantities like QCD factor, renormalisation group invariant parameter are same for both SM and nmUED scenarios.}
\be
\frac{(\Delta M_s)_{\rm nmUED}}{(\Delta M_s)_{\rm SM}}=
\frac{S(x_t,r_f,r_V,R^{-1})}{S_0(x_t)}~.
\label{delmsnmued}
\ee
Let us focus on the above Eq.\;\ref{delmsnmued}. It is clear from the Eq.\;\ref{SACD} that the function $S(x_t,r_f,r_V,R^{-1})$ (numerator of the right hand side of the Eq.\;\ref{delmsnmued}) represents the combine contributions of SM and the NP (in the present case KK-contributions) obtained from the the box diagram given in Fig.\;\ref{bb_box}. Besides, the function $S_0(x_t)$ (denominator of the right hand side of the Eq.\;\ref{delmsnmued}) represents the SM contributions only. Moreover, from the above Eq.\;\ref{delmsnmued} it is obvious that the ratio of the function $S(x_t,r_f,r_V,R^{-1})$ to the function $S_0(x_t)$ is equivalent to $\frac{(\Delta M_s)_{\rm nmUED}}{(\Delta M_s)_{\rm SM}}$. Therefore, from this ratio we have the opportunity to estimate the effects of BLT parameters ($r_V$, $r_f$) and $R^{-1}$ by the virtue of $\Delta B=2$ transitions. We can compare this ratio\footnote{Here, we would like to mention that, in order to measure the effect of Georgi-Machacek model on $\Delta M_s$, the same procedure has been adopted in ref. \cite{Banerjee:2019gmr}. } with the quantity $\frac{(\Delta M_s)_{\rm exp}}{(\Delta M_s)_{\rm SM}}$. Now the value of SM prediction for $\Delta M_s$ is $(18.77\pm 0.86){\rm ps}^{-1}$ i.e., $(\Delta M_s)_{\rm SM}=(18.77\pm 0.86){\rm ps}^{-1}$ \cite{Lenz:2019lvd}. On the other hand experimental value for the same is $(17.749\pm 0.019\pm 0.007){\rm ps}^{-1}$, i.e., $(\Delta M_s)_{\rm exp}=(17.749\pm 0.019\pm 0.007){\rm ps}^{-1}$ \cite{Zyla:2020zbs}. Note that in both cases the errors are given with 1$\sigma$ uncertainty. Now using these values we can easily determine the value\footnote{Since in this case $(\Delta M_s)_{\rm SM}$ and $(\Delta M_s)_{\rm exp}$ are independent quantities, therefore the uncertainty of the ratio $\frac{(\Delta M_s)_{\rm exp}}{(\Delta M_s)_{\rm SM}}$ has been determined from the quadrature sum method. For the purpose of determination of the value of this quantity with uncertainty, we have followed the method given in the book, {\it An introduction to Error Analysis}, by John R. Taylor.} of the ratio $\frac{(\Delta M_s)_{\rm exp}}{(\Delta M_s)_{\rm SM}}$ and the corresponding value is $(0.946\pm 0.043)$, i.e., $\frac{(\Delta M_s)_{\rm exp}}{(\Delta M_s)_{\rm SM}}=(0.946\pm 0.043)$. In this case also the error is given with 1$\sigma$ uncertainty. However, while we perform the numerical analysis, we compare the quantity $\frac{(\Delta M_s)_{\rm nmUED}}{(\Delta M_s)_{\rm SM}}$ with the 2$\sigma$ allowed range of the obtained value of this ratio, i.e., $(0.946\pm 0.086)$. Variation of this quantity with respect to $R^{-1}$ is given in Fig.\;\ref{fig2e}. With this treatment, we have estimated the effects of $R^{-1}$ and BLT parameters ($r_V$, $r_f$) on the $\Delta M_s$ in nmUED scenario.  We can also obtain lower limit on $R^{-1}$ form this ratio $\frac{(\Delta M_s)_{\rm nmUED}}{(\Delta M_s)_{\rm SM}}$.
}

\section{Numerical analysis}\label{anls}
In the current article, for the first time we have computed the KK-contributions to the WC of the operator
($\Delta B = 2$) in the nmUED scenario. The function $S_n(x_t,x_{f^{(n)}},x_{V^{(n)}})$ given in Eq.~\ref{SFIN} corresponds to the $n^{th}$ level KK-contributions to the coefficient for the $\Delta B = 2$ operator. The function $S_n$ contains the dependence of KK-masses of gauge boson as well as fermion in the nmUED scenario. Besides, in view of the analysis of the effect of the SM Higgs mass on vacuum stability in UED model \cite{Datta:2012db}, we consider the sum of KK-contributions up to 5 KK-levels\footnote{ In earlier studies, usually 20-30 KK-levels have been taken while adding up the contributions from KK-modes.} and thereafter we add up the total KK-contributions with the SM counterpart\footnote{We have taken $M_W=80.379$ GeV \cite{Zyla:2020zbs} for SM $W^\pm$ boson mass and $\overline{m}_t(m_t)=165.25$ GeV \cite{CKM:summer} for SM top quark mass.}. Moreover, due to the converging\footnote{The summation of KK-contribution is convergent in UED type models with one extra space-like dimension, as far as one-loop calculation is concerned\cite{Dey:2004gb}.} nature of KK-summation, the numerical values would not differ drastically whether one considers higher numbers of KK-levels during the evaluation of KK-contributions for the loop diagrams\cite{Datta:2015aka, Datta:2016flx, Shaw:2019fin}. 

\subsection{Possible constraints and range of BLT parameters}
Here we highlight the following constraints that have been considered in our analysis.
\begin{itemize}
\item 
In this nmUED scenario, comprehensive analyses on different rare $B$-decay processes (FCNC type), for example $B_s\rightarrow \mu^+\mu^-$ \cite{Datta:2015aka}, $B\rightarrow X_s\gamma$ \cite{Datta:2016flx} and $B\rightarrow X_s\ell^+\ell^-$ \cite{Shaw:2019fin} have been performed. Moreover, these processes have always been played significant role for searching any favourable kind of NP scenario. In all these cases the expressions of these observables are the functions of the same set of parameters i.e., $r_V$, $r_f$ and $R^{-1}$ which are also involved in the expressions of the observables of the current article. Therefore, in this article it is very necessary that we should deal with such parameter space which is satisfied by the experimental data of these observables. Using the expressions of ${\rm Br}(B_s\rightarrow \mu^+\mu^-)$, ${\rm Br}(B\rightarrow X_s\gamma)$ and ${\rm Br}(B\rightarrow X_{_s}\ell^+\ell^-)$ given in \cite{Datta:2015aka}, \cite{Datta:2016flx} and \cite{Shaw:2019fin} we have considered the branching ratios of these rare decay processes as constraints in our present study. In the following we present the latest experimental data for branching ratios of these processes 
\begin{table}[H]
\begin{center}
\begin{tabular}{|c|c|c|}
\hline
\multicolumn{2}{|c|}{{Observables}}  & {Experimental value}   \\ \hline
\multicolumn{2}{|c|}{${\rm Br}(B_s\rightarrow \mu^+\mu^-)$} & {$(3.1\pm 0.6)\times 10^{-9}$} \cite{Amhis:2019ckw} \\
\hline
\multicolumn{2}{|c|}{${\rm Br}(B\rightarrow X_s\gamma)$} & {$(3.32\pm 0.15)\times 10^{-4}$} \cite{Amhis:2019ckw} \\
\hline
\multirow{2}{*}{${\rm Br}(B\rightarrow X_{_s}\ell^+\ell^-)$} & $q^2 \in [1,6] {\rm GeV}^2$ & {$(1.60^{+0.41+0.17}_{-0.39-0.13}\pm 0.18)\times 10^{-6}$} \cite{Lees:2013nxa} \\ \cline{2-3}
& $q^2 \in [14.4,25] {\rm GeV}^2$ & {$(0.57^{+0.16+0.03}_{-0.15-0.02}\pm 0.00)\times 10^{-6}
$} \cite{Lees:2013nxa} \\ \hline
\end{tabular}
\caption{Experimental data for branching ratios of $B_{s} \rightarrow \mu^+ \mu^-$, $B\rightarrow X_s\gamma$ and $B\rightarrow X_{_s}\ell^+\ell^-$.}
\label{t:4}
\end{center}
\end{table}
\item Electroweak precision test (EWPT) is significant and instrumental for constraining any kind of BSM physics. Using the technique of the correction to Fermi constant $G_F$ at tree level one can perform the corrections to Peskin-Takeuchi parameters S, T, and U in the nmUED model. This is a distinctive feature of this nmUED scenario with respect to the minimal version of the UED model where these corrections emerge at one-loop processes. A detailed exercise on EWPT in this nmUED model has been given in \cite{Datta:2015aka, Biswas:2017vhc}. The S, T, and U parameters in the nmUED model are the functions of $r_V$, $r_f$ and $R^{-1}$. Following the similar approach provided in refs. \cite{Datta:2015aka, Biswas:2017vhc} we have imposed EWPT as one of the constraints in the analysis.  

\end{itemize}
To this end, we would like to discuss the range of values of BLT parameters used in our present exercise. In general values of the BLT parameters may be negative or positive. However, it is clearly seen from Eq.\;\ref{norm} that, for ${r_f}/{R}=-\pi$ the zero-mode solution becomes divergent and beyond  ${r_f}/{R} = - \pi$ the zero-mode fields become ghost-like. Therefore, any values of BLT parameters lower than $- \pi R$ should not be considered, but for the purpose of completeness we have taken some negative values of BLT parameters in our numerical analysis. However, from the study of electroweak precision data \cite{Datta:2015aka, Biswas:2017vhc} large portion of negative values of BLT parameters have been disfavoured.

\subsection{Results}
{At this stage, considering the above mentioned constraints, we would like to find the parameter space which satisfy the 2$\sigma$ allowed ranges of the CKM elements and $\frac{(\Delta M_s)_{\rm exp}}{(\Delta M_s)_{\rm SM}}$ simultaneously. In the present version of nmUED scenario we have three independent free parameters e.g., inverse of radius of compactification $R^{-1}$, dimensionless scaled BLT parameters for boson $R_V(=r_V/R)$ and fermion $R_f(=r_f/R)$. In order to find the allowed parameter space we have chosen the following ranges for the free parameters:
\begin{equation}
R^{-1}\in [0.05, 2] {\rm TeV}\;\; ; R_V\in [-3,20]\;\; {\rm and}\;\; R_f\in [-3,20]\;\;.
\label{range}
\end{equation}

Using the above ranges of free parameters we have obtained a region in $R_V-R_f$ plane shown in Fig.\;\ref{parame}. Moreover, for different combination of $R_V, R_f$ the allowed values of $R^{-1}$ are displayed by colour codes. The Fig.\;\ref{parame} shows that, with the increasing values of $R_V$ and  $R_f$ the allowed values of $R^{-1}$ are increased. This can be explained in the following way. Since with the increasing values of BLT parameters the KK-masses are decreased, consequently with the decreasing values of KK-masses the loop function $S$ (obtained from one-loop box diagrams) is enhanced. Therefore, in order to compensate this enhancement one requires the increasing values of $R^{-1}$ (as KK-masses are increased with the increasing values of $R^{-1}$). Another notable feature is that for $R_V\approx R_f$ most of the allowed values of $R^{-1}$ with higher in magnitude are appeared and these type of points are increased with the larger values of BLT parameters. Moreover, it is evident from the figure that only a few portion of negative values of BLT parameters are allowed. The reason is that, apart from the constraints on branching ratio of several rare $B$-decay process we have considered EWPT as crucial one. As a consequences, in this nmUED scenario EWPT favours the region where $R_V\approx R_f$ and larger values of $R^{-1}$ are disfavoured for most of the negative values of BLT parameters \cite{Datta:2015aka, Biswas:2017vhc}. Furthermore, we would like to mention that, among the different constraints of $B$-physics observables (Br($B_s\to \mu^+\mu^-$) \cite{Datta:2015aka}, Br($B\to X_s\gamma$) \cite{Datta:2016flx} and Br($B\to X_s\ell^+\ell^-$) \cite{Shaw:2019fin}) Br($B_s\to \mu^+\mu^-$) \cite{Datta:2015aka} is the most dominating in nature. Therefore, as far as the $B$-physics is concerned, before the present analysis of $\Delta B=2$ transitions of this article, the most admissible value of lower limit of $R^{-1}$ was derived from the Br($B_s\to \mu^+\mu^-$) \cite{Datta:2015aka} for a specific combination of BLT parameter in the present version of nmUED scenario. However, from the current analysis of $\Delta B=2$ transitions, we have found a distinguishable observation with respect to our previous analyses on rare $B$-decay process \cite{Datta:2015aka, Datta:2016flx, Shaw:2019fin}. For example, in contrast to the previous analyses \cite{Datta:2015aka, Datta:2016flx, Shaw:2019fin}, here we obtain lower limit of $R^{-1}$ with larger values for several combination of positive values of scaled BLT parameters ($R_V, R_f$). In the following, for the purpose of illustration, we have picked up two sets of nonvanishing scaled BLT parameters ($R_V, R_f$) from the allowed region shown in the Fig.\;\ref{parame} and consequently using the Fig.\;\ref{BP1} we will discuss the characteristic dependence of the observables which have been considered in this article with respect to $R^{-1}$. Moreover, depending on the BLT parameters we can estimate the lower bound on the $R^{-1}$ (using the Fig.\;\ref{BP1}) that are obtained from the present analysis of $\Delta B=2$ transition. 

\begin{figure}[t!]
\begin{center}
\includegraphics[height=10cm,width=14cm,angle=0]{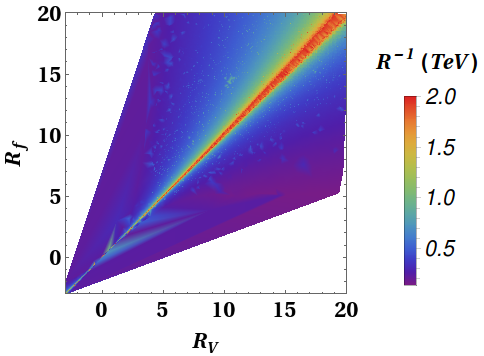}
\caption{Allowed parameter space in $R_V(=r_V/R)-R_f(=r_f/R)$ plane satisfying 2$\sigma$ range of latest values of CKM elements and $\frac{(\Delta M_s)}{(\Delta M_s)_{\rm SM}}$ simultaneously. Moreover, the allowed values of $R^{-1}$ for different combination of $R_V(=r_V/R), R_f(=r_f/R)$ are indicated by colour codes. Here we sum the contributions up to 5 KK-levels in loop function $S$ while calculating box diagram.}
\label{parame}
\end{center}
\end{figure}

The Fig.\;\ref{BP1} contains 5 panels that have shown the dependence of the variables $|V_{td}|$, $\gamma$ (in degree), $\bar{\eta}$, $\bar{\rho}$ and $\frac{(\Delta M_s)}{(\Delta M_s)_{\rm SM}}$ with respect to $R^{-1}$. In each panel we have shown the dependence of these observables for three benchmark points which have been chosen from the allowed parameter space shown in Fig.\;\ref{parame}. The first benchmark point (BP1) is indicated by the red coloured solid line and for this case $R_V=6.72$ and $R_f=6.58$. The second benchmark point (BP2) is indicated by the blue coloured dashed line and for this case $R_V=8.76$ and $R_f=9.06$. Finally, in order to show the UED results we have taken another point (also resides within the region shown in Fig.\;\ref{parame}) which is indicated by black coloured solid line for which $R_V=0$ and $R_f=0$. 

Let us focus on the BP1. For this benchmark point, when the curve of each observables intersects the 2$\sigma$ allowed range of the corresponding observables, then we obtain the values of lower limits of $R^{-1}$ for the observables. These values of lower limits are given in table \ref{lowerlimRin} in GeV unit. From this table it is clear that, for BP1 the highest value of lower limit of $R^{-1}$ is obtained from the observable $\gamma$ (e.g., 952.09 GeV) while the lowest value of lower limit is generated from the observable $\bar{\eta}$ (e.g., 888.32 GeV). Let us now discuss the nature of these curves. First of all, using the Eq.\;\ref{DMQ} we have derived the value of $|V_{td}|$ and it is governed by the function $S$ which is dependent on the model parameters $R_V, R_f$ and $R^{-1}$. Now for a fixed value of $R_V$ and $R_f$ the KK-masses are controlled by $R^{-1}$. With the increasing values of $R^{-1}$ the KK-masses are increased, hence the loop function $S$ is decreased. Therefore, naturally, the value of $|V_{td}|$ is increased with $R^{-1}$. However, after a certain value of $R^{-1}$, when the KK-masses are very high then decoupling behavior of the KK-mode contribution arises and consequently there is no variation of $|V_{td}|$ with $R^{-1}$. After determination of $|V_{td}|$ we have derived $R_t$ using Eq.\;\ref{2.95}. Thereafter, with this $R_t$ we have derived $\bar{\rho}$, $\bar{\eta}$ and $\gamma$ from the Eq.\;\ref{rhoetagama}. It is evident from Eq.\;\ref{2.95}, that for the given values of $\lambda$ and $V_{cb}$, $R_t$ is proportional to $|V_{td}|$, therefore $R_t$ will follow the same characteristics as $|V_{td}|$ with $R^{-1}$. Moreover, it is also evident from the Eq.\;\ref{rhoetagama} that $\bar{\eta}$ is proportional to $R_t$ (for a given value of $\beta$), therefore, $\bar{\eta}$  will be increased with the increasing values of $R^{-1}$ and obviously $\bar{\rho}$ will be decreased. Further, the increasing behaviour of $\gamma$ with respect to $R^{-1}$ is a resultant factor of increment of $\bar{\eta}$ (with $R^{-1}$) and decrement of $\bar{\rho}$ (with $R^{-1}$). At the KK-mode decoupling limit, the values of all observables are saturated and do not show any variation with $R^{-1}$. Now the behaviour of the curve given in the last panel (\ref{fig2e}) of the Fig.\;\ref{BP1} can be explained in the following way. As we have already mentioned that the nmUED scenario is a class of MFV models, therefore, the NP contribution to one-loop box diagram is always positive. Hence, $\frac{(\Delta M_s)}{(\Delta M_s)_{\rm SM}}$ \bigg(for the sake of notational simplicity here we assume, $(\Delta M_s)\equiv (\Delta M_s)_{\rm nmUED}$, while $(\Delta M_s)=(\Delta M_s)_{\rm SM}+ (\Delta M_s)_{\rm NP}$\bigg) is greater than one. Further, with the increasing values of $R^{-1}$ (basically KK-mass) the one-loop function $S$ is decreased, therefore, the quantity $\frac{(\Delta M_s)}{(\Delta M_s)_{\rm SM}}$ is decreased and after a certain large value of $R^{-1}$ it will become one.

\begin{figure}[htbp!]
\centering
\vspace*{-1cm}
\subfloat[]{\label{fig2a}\includegraphics[height=5.5cm,width=9cm,angle=0]{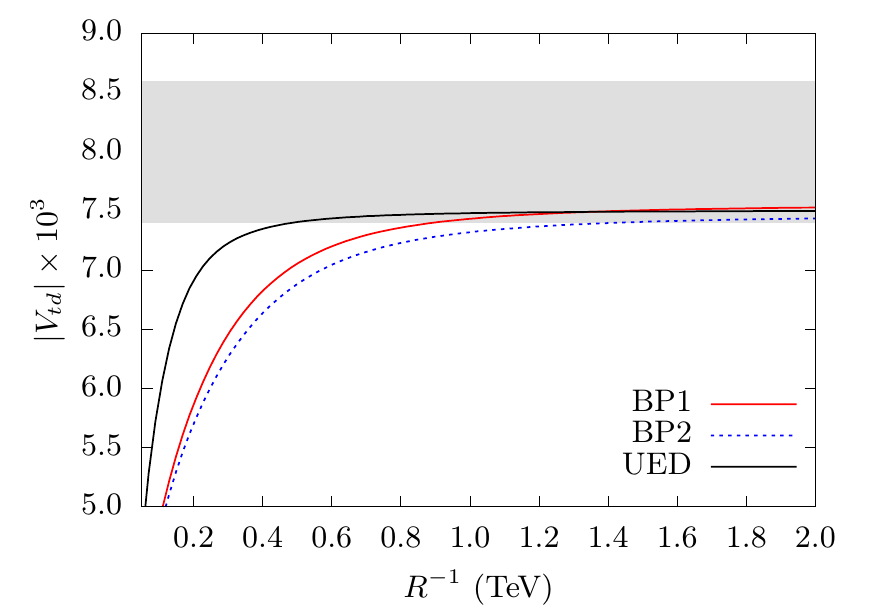}}
\subfloat[]{\label{fig2b}\includegraphics[height=5.5cm,width=9cm,angle=0]{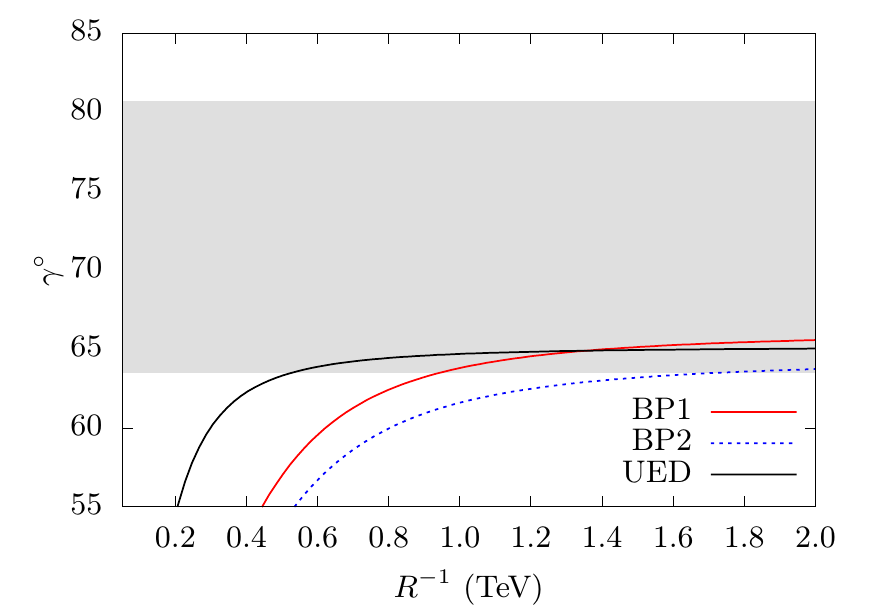}}\\
\subfloat[]{\label{fig2c}\includegraphics[height=5.5cm,width=9cm,angle=0]{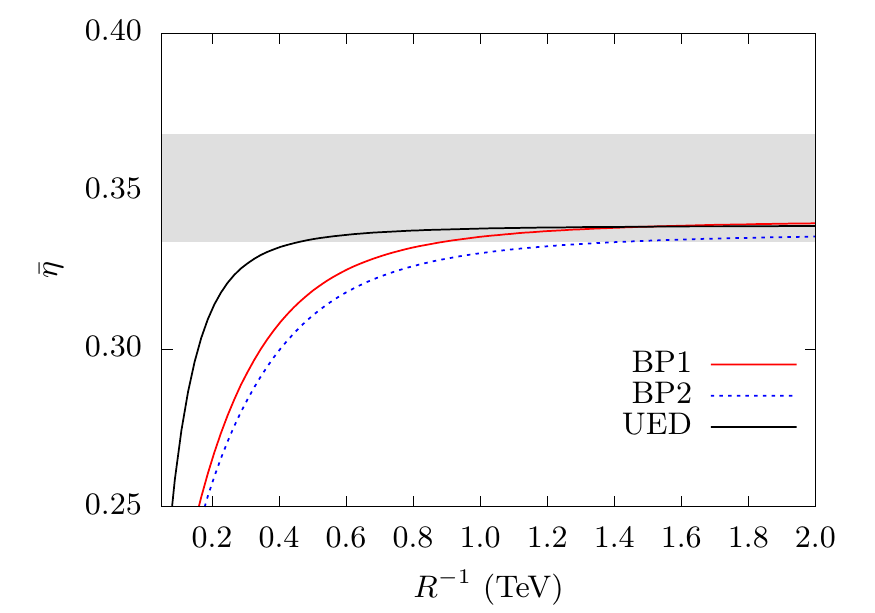}}
\subfloat[]{\label{fig2d}\includegraphics[height=5.5cm,width=9cm,angle=0]{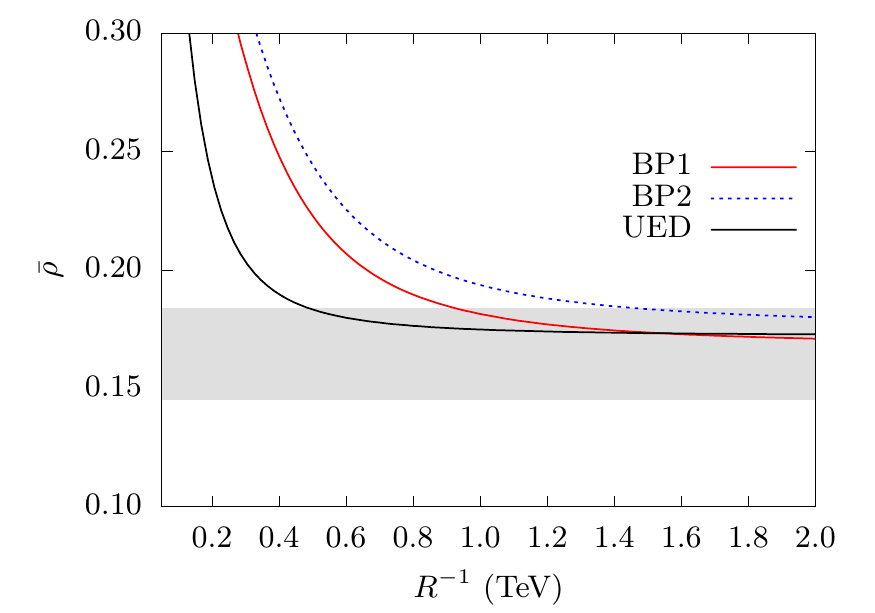}}\\
\subfloat[]{\label{fig2e}\includegraphics[height=5.5cm,width=9cm,angle=0]{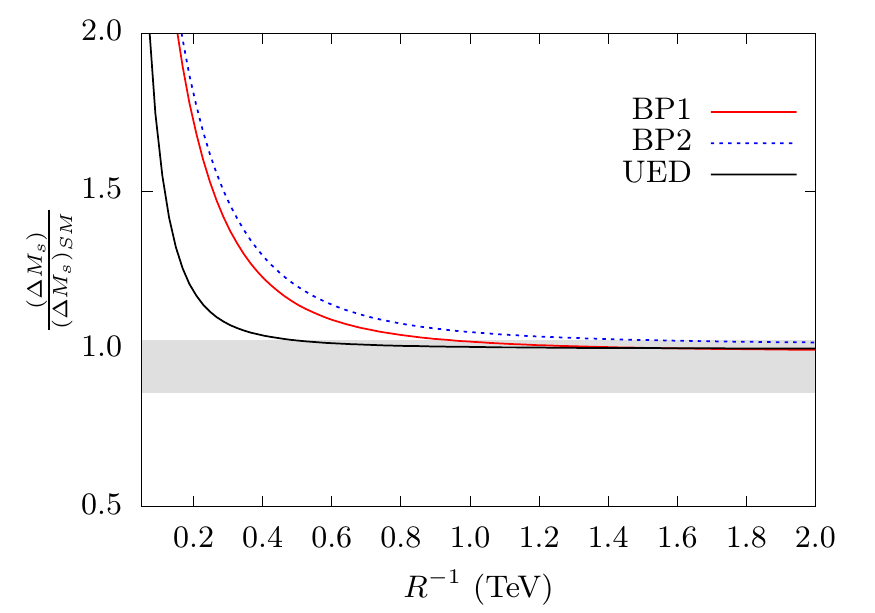}}
\caption{Variation of CKM parameters [(a):$|V_{td}|$; (b):$\gamma^\circ$; (c):$\bar{\eta}$; (d):$\bar{\rho}$] and (e):$\frac{(\Delta M_s)}{(\Delta M_s)_{\rm SM}}$ with respect to $R^{-1}$, where we have chosen three sets of scaled BLT parameters $R_V(=r_V/R)=6.72, R_f(=r_f/R)=6.58$ (considered as BP1 indicated by solid red line); $R_V(=r_V/R)=8.76, R_f(=r_f/R)=9.06$ (considered as BP2 indicated by doted blue line) and $R_V(=r_V/R)=0, R_f(=r_f/R)=0$ (considered as UED indicated by solid black line) from the allowed parameter space as shown in the Fig.\;\ref{parame}. Here we sum the contributions up to 5 KK-levels in loop function $S$ while calculating box diagrams. The horizontal gray band shows the 2$\sigma$ allowed range of respective observables.}
\label{BP1}
\end{figure}

All the above mentioned explanations for all the observables also hold good for BP2. However, for BP2 the lower limits on $R^{-1}$ are changed and they are also presented in table \ref{lowerlimRin} in GeV unit. From the values of lower limits of $R^{-1}$ (that are given in table \ref{lowerlimRin} for BP2) it is clear that the lower limits of $R^{-1}$ for BP2 are larger. As for example the highest value (which is obtained from $\gamma$) is 1482.38 GeV and the lowest value (which is obtained from $\bar{\eta}$) is 1329.05 GeV. This enhancement of lower limit with respect to the BP1 can easily be explained in the following way. Since in the case of BP2, the values of the BLT parameter are higher than that of the BP1, therefore the KK-masses are decreased. Hence, in order to compensate the KK-mass decrement larger value of $R^{-1}$ is required. Therefore, for BP2 we obtain larger values of lower limits on $R^{-1}$ for all observables.

\begin{table}[htbp!]
\begin{center}
\begin{tabular}{|c|c|c|c|c|c|} 
\hline 
{} & from $|V_{td}|$ & from $\gamma^\circ$ & from $\bar{\eta}$  &  from $\bar{\rho}$ & from $\frac{(\Delta M_s)}{(\Delta M_s)_{\rm SM}}$ \\
\hline 
BP1  & 898.56 & 952.09  &  888.32  & 922.74 &  895.09 \\
\hline
BP2  & 1381.58  &  1482.38 &  1329.05  & 1418.51 & 1335.04 \\
\hline\hline
UED  & 482.20 & 535.48 &  461.41  &  488.96 & 469.95 \\
\hline
\end{tabular}
\caption{Lower limits on $R^{-1}$ (in GeV) evaluated for various CKM parameters [$|V_{td}|$, $\gamma$ (in degree), $\bar{\eta}$, $\bar{\rho}$] and $\frac{(\Delta M_s)}{(\Delta M_s)_{\rm SM}}$ for the selected benchmark points (BP1 and BP2) and UED. For each panel in Fig.\;\ref{BP1}, the lower limits (on $R^{-1}$) have been obtained from intersection of the respective curves with the 2$\sigma$ allowed range of the corresponding observables.}
\label{lowerlimRin}
\end{center}
\end{table}

At this stage we would like to emphasize that the current analysis of $\Delta B=2$ transitions provide us better result with respect to our earlier analyses \cite{Datta:2015aka, Datta:2016flx, Shaw:2019fin} on rare decays of $B$-meson in the same version of nmUED scenario. From the previous discussions it is already clear that we can push the lower limit on $R^{-1}$ to appreciable higher range ($\approx 1.48$ TeV or even higher) for favourable choice BLT parameters. On the other hand, for illustrative purpose we have explicitly checked the lower limits of $R^{-1}$ for two benchmark points (BP1 and BP2) for Br$(B_s\to \mu^+\mu^-)$. The values of the lower limits are lesser than that obtained from the current analysis of $\Delta B=2$ transitions. For example for BP1 the lower limit of $R^{-1}$ from Br$(B_s\to \mu^+\mu^-)$ is 841 GeV while for BP2 the corresponding value is 1.26 TeV\footnote{Here, among the all $B$-physics constraints that have been considered in the present article,  we have mentioned the lower limit of $R^{-1}$ for two benchmark points for Br$(B_s\to \mu^+\mu^-)$ only. Before the present analysis of $\Delta B=2$ transitions, the most dominant constraint came from Br$(B_s\to \mu^+\mu^-)$ in nmUED scenario.}. Hence, comparing these two values of lower limit of $R^{-1}$ with the same obtained from the analysis of $\Delta B=2$ transition (e.g., 952.09 GeV for BP1  and 1.48 TeV for BP2) we can readily infer that we have obtained so far the most dominating constraining power from our present analysis of $\Delta B=2$ transitions. Therefore, from $B$-physics perspective, these lower limits, that we have obtained from the present analysis of $\Delta B=2$ transitions, are so far the most admissible results in the current version of nmUED scenario.

Before we conclude, we would like to remark on the lower limits on $R^{-1}$ which are achieved in the UED scenario considering the current analysis on $\Delta B=2$ transitions. We can achieve the UED results from our analysis when BLT parameters vanish, i.e., for $R_V=R_f=0$. In this set up KK-mass for $n^{th}$ KK-level simply emerges as $nR^{-1}$. On the other hand, the overlap integrals $I^n_1$ and $I^n_2$ become unity. Hence, with this limit, the function $S_n(x_t,x_{f^{(n)}},x_{V^{(n)}})$ is converted into its UED form. We have independently checked that when the BLT parameters are zero the expressions of the function $S_n(x_t,x_{f^{(n)}},x_{V^{(n)}})$ is exactly matched with that of the given in ref. \cite{Buras:2002ej}\footnote{In the article \cite{Buras:2002ej}, the authors have not considered any radiative corrections to the KK-masses in their analysis. Therefore, the KK-mass at the $n^{th}$ KK-mode is $nR^{-1}$.}. Like the chosen benchmark points BP1 and BP2 the values of lower limits of $R^{-1}$ for $R_V = R_f =0$ are also presented in table \ref{lowerlimRin}. In this case the highest value (which is obtained from $\gamma$) is 535.48 GeV and the lowest value (which is obtained from $\bar{\eta}$) is 461.41 GeV. Considering $B$-physics analyses, these values of lower limits of $R^{-1}$ are slightly improved, because the value of these lower limits are slightly higher than that obtained from our previous analyses \cite{Datta:2015aka, Datta:2016flx, Shaw:2019fin}. However, it is needless to say that, these limits are not very striking values, but close to those values that have been achieved from previous analyses in UED scenario. As for example $(g-2)_\mu$ \cite{Nath:1999aa}, $\rho$-parameter \cite{Appelquist:2002wb}, FCNC process \cite{Buras:2003mk,Buras:2002ej, Agashe:2001xt, Chakraverty:2002qk}, $Zb\bar{b}$ \cite{Jha:2014faa, Oliver:2002up} and electroweak observables \cite{Strumia:1999jm, Rizzo:1999br, Carone:1999nz} provide a lower bound of about 300-600 GeV on $R^{-1}$. Besides, from the analysis of projected tri-lepton signal at 8 TeV LHC one can  obtain lower limit on $R^{-1}$ up to 1.2 TeV\cite{Belyaev:2012ai, Golling:2016gvc, Gershtein:2013iqa}. At this stage it is needed  to mention that the value of lower limits on $R^{-1}$ from the analysis on $\Delta B=2$ transitions for minimal version of UED scenario, have already been excluded by the LHC data. Since the recent analyses including LHC data have ruled out $R^{-1}$ up to 1.4~TeV~\cite{Choudhury:2016tff, Beuria:2017jez, Chakraborty:2017kjq, Deutschmann:2017bth}.

\section{Summary}\label{concl}
In this article we estimate the Kaluza-Klein contribution to the $\Delta B=2$ transitions in a class of (4+1)-dimensional Universal Extra Dimensional (in which all Standard Model particle can propagate along the extra spatial dimension) scenario in the presence of boundary localised terms (BLTs). The coefficient of these terms are parametrised to the unknown radiative corrections for the masses and couplings of Kaluza-Klein modes. Due to the presence of these boundary terms the masses and coupling strengths are nontrivially modified in 4-dimensional effective theory with respect to the minimal version of the Universal Extra Dimensional scenario. Utilising two different kinds of BLT parameters e.g., $r_V$ (represents the coefficients of boundary terms for the gauge and Higgs sectors) and $r_f$ (specifies the coefficients of boundary terms of fermions and Yukawa interactions) we have investigated the $\Delta B=2$ transitions in nonminimal Universal Extra Dimensional scenario.

The effective Hamiltonian for the $\Delta B=2$ transitions can be expressed by four-fermion interactions and the coefficient of the interactions are parametrised by appropriate Wilson Coefficient.
With the computation of one-loop box diagrams given in Fig.\;\ref{bb_box} we have evaluated the coefficient for the operator that is responsible for the $\Delta B=2$ transitions. Moreover, utilising the Glashow Iliopoulos Maiani mechanism we have included contributions from three generations of quarks in our analysis. On the other hand, considering a recent analysis relating the Higgs boson mass and cut-off of a Universal Extra Dimensional theory  \cite{Datta:2012db} we summed up to five Kaluza-Klein modes in our computation. Further, in view of the fact that the nonminimal Universal Extra Dimensional scenario belongs to a class of Minimal Flavour Violation models, we have simply added the Kaluza-Klein contributions coming from the one-loop box diagrams (Fig.\;\ref{bb_box}) to the corresponding Standard Model (zero-mode) contributions. 

After evaluation of the function $S$ (obtained from one-loop box diagrams) we have determined several elements e.g., $|V_{td}|, \bar{\eta}, \bar{\rho}$ and $\gamma$ in nonminimal Universal Extra Dimensional scenario. Some of these quantities have played very important role for Wolfenstein parametrisation and further using these quantities we can estimate geometrical shape of unitarity triangle. Finally, we have evaluated the quantity $\Delta M_s$ scaled by the corresponding Standard Model value. Comparing our theoretical predictions of these quantities with the corresponding 2$\sigma$ allowed ranges, we have constrained the parameter space of the present version of nonminimal Universal Extra Dimensional scenario. Moreover, in our analysis we have considered the branching ratios of some important rare decay processes of $B$-meson: such as $B_s\to \mu^+\mu^-$, $B\to X_s\gamma$ and $B\to X_s\ell^+\ell^-$ as well as electroweak precision data as constraints. 

It has already been alluded that for the vanishing BLT parameters (i.e., $R_f = R_V = 0$) we can reproduce the results of the minimal version of Universal Extra Dimensional scenario. Therefore, using our analysis, we have reexamined the lower limit on $R^{-1}$ in the framework of minimal Universal Extra Dimensional scenario in the vanishing BLT limit. In that case the value of the lower limit on $R^{-1}$ becomes 535.48 GeV (highest value that we have obtained from our present analysis of $\Delta B=2$ transition for UED) and it is slightly higher than those values that are determined from our earlier studies on $B$-physics. However, this value is excluded from recent collider analysis at the LHC.

Nevertheless, in the presence of different nonvanishing BLT parameters we can enrich the results of lower limit on $R^{-1}$ in the current version of nonminimal Universal Extra Dimensional scenario. As for example if we choose a set of BLT parameter ($R_V=6.72$ and $R_f=6.58$, i.e., for BP1) from the allowed parameter space given in Fig.\;\ref{parame} then using our analysis we can obtain the lower limit of $R^{-1}$ $\approx 952$ GeV (highest value that we have obtained from our present analysis of $\Delta B=2$ transition for BP1). Moreover, the same can be even higher (e.g., $\approx 1.48$ TeV) if we choose another set of BLT parameter ($R_V=8.76$ and $R_f=9.06$, i.e., for BP2) from the allowed parameter space shown in Fig.\;\ref{parame}. Definitely, these results (lower limits on $R^{-1}$) in the current version of nonminimal Universal Extra Dimensional scenario are the most admissible values in comparison to the limits obtained from our earlier analyses on rare decays of $B$-meson \cite{Datta:2015aka, Datta:2016flx, Shaw:2019fin}. 
}

{\bf Acknowledgements} The author would like to give thanks to Anirban Biswas for computational support. The author thanks Anindya Datta for some illuminating suggestions.
\begin{appendices}
\renewcommand{\thesection}{\Alph{section}}
\renewcommand{\theequation}{\thesection-\arabic{equation}} 

\setcounter{equation}{0}  
\section{Feynman rules required for the study of \boldmath{$\Delta B=2$} transitions in nmUED}\label{fyerul}
This Appendix contains the Feynman rules required for our calculations with the assumption that all momenta and fields are incoming:

1) $G^{\mu}{\overline{f}_1} f_2$
  $\displaystyle  : {i g_s}{T^a_{\alpha\beta}} \gamma_\mu C$, where $C$ takes the following form:

\begin{equation}
 \begin{aligned}
  G^{\mu} \bar{u_i} u_i: C &= 1,\\   
  G^{\mu} {\overline{T}^{1(n)}_i} T^{1(n)}_i: C &= 1,\\
  G^{\mu} {\overline{T}^{2(n)}_i} T^{2(n)}_i: C &= 1,\\ 
  G^{\mu} {\overline{T}^{1(n)}_i} T^{2(n)}_i: C &= 0,\\
  G^{\mu} {\overline{T}^{2(n)}_i} T^{2(n)}_i: C &= 0.
\end{aligned}
\end{equation}
2) $S^{\pm}{\overline{f}_1} f_2$
  $\displaystyle  = \frac{g_2}{\sqrt{2} M_{W^{(n)}}} (P_L C_L + P_R C_R)$, where $C_L$ and $C_R$ are expressed in the following way:

\begin{equation}
\begin{aligned}
  & G^+ \bar{u_i} d_j :  &
  &\left\{\begin{array}{l}C_L = -m_i V_{ij},\\
      C_R = m_j V_{ij},\end{array}\right.
  &&G^- \bar{d_j} u_i :     &
  &\left\{\begin{array}{l}C_L = -m_j V_{ij}^*,\\
      C_R = m_i V_{ij}^*,\end{array}\right.\\
  & G^{(n)+}{\overline{T}^{1(n)}_i} d_j :  &
  &\left\{\begin{array}{l}C_L = -m_1^{(i)} V_{ij},\\
      C_R = M_1^{(i,j)} V_{ij},\end{array}\right.
  &&G^{(n)-}\bar{d_j}T^{1(n)}_i :   &
  &\left\{\begin{array}{l}C_L = -M_1^{(i,j)} V_{ij}^*,\\
     C_R = m_1^{(i)} V_{ij}^*,\end{array}\right.\\
  & G^{(n)+}{\overline{T}^{2(n)}_i} d_j :  &
  &\left\{\begin{array}{l}C_L = m_2^{(i)} V_{ij},\\
      C_R =-M_2^{(i,j)} V_{ij},\end{array}\right.
  &&G^{(n)-}\bar{d_j}T^{2(n)}_i :   &
  &\left\{\begin{array}{l}C_L = M_2^{(i,j)} V_{ij}^*,\\
     C_R =-m_2^{(i)} V_{ij}^*,\end{array}\right.\\
  & H^{(n)+}{\overline{T}^{1(n)}_i} d_j :  &
  &\left\{\begin{array}{l}C_L = -m_3^{(i)} V_{ij},\\
      C_R = M_3^{(i,j)} V_{ij},\end{array}\right.
  &&H^{(n)-}\bar{d_j}T^{1(n)}_i :   &
  &\left\{\begin{array}{l}C_L = -M_3^{(i,j)} V_{ij}^*,\\
     C_R = m_3^{(i)} V_{ij}^*,\end{array}\right.\\
  & H^{(n)+}{\overline{T}^{2(n)}_i} d_j :  &
  &\left\{\begin{array}{l}C_L = m_4^{(i)} V_{ij},\\
      C_R =-M_4^{(i,j)} V_{ij},\end{array}\right.
  &&H^{(n)-}\bar{d_j}T^{2(n)}_i :   &
  &\left\{\begin{array}{l}C_L = M_4^{(i,j)} V_{ij}^*,\\
     C_R =-m_4^{(i)} V_{ij}^*.\end{array}\right.
\end{aligned}
\end{equation}

3) $W^{\mu\pm}{\overline{f}_1}f_2$
  $\displaystyle  :  \frac{i g_2}{\sqrt{2}} \gamma_\mu P_L C_L$, where $C_L$ takes the following form \cite{Datta:2015aka}:
\begin{equation}
\begin{aligned}
  & W^{\mu+}\bar{u_i} d_j : &&     C_L = V_{ij},
  && W^{\mu-}\bar{d_j} u_i : &&    C_L = V^*_{ij},\\
  & W^{\mu(n)+}{\overline{T}^{1(n)}_i}d_j : &&   C_L = I^n_1\;c_{in} V_{ij},
  &&W^{\mu(n)-}\bar{d_j}{{T}^{1(n)}_i} : && C_L = I^n_1\;c_{in} V^*_{ij},\\
  & W^{\mu(n)+}{\overline{T}^{2(n)}_i}d_j : &&   C_L = -I^n_1\;s_{in} V_{ij},
  &&W^{\mu(n)-}\bar{d_j}{{T}^{2(n)}_i} : && C_L = -I^n_1\;s_{in}V^*_{ij},
\end{aligned}
\end{equation}
where the fermion fields $f\equiv u, d, T^1_t, T^2_t$.

The mass parameters $m_x^{(i)}$ are expressed in the following way \cite{Datta:2015aka}:
\begin{equation}
\label{mparameters}
  \begin{aligned}
    m_1^{(i)} &= I^n_2\;m_{V^{(n)}}c_{in} +I^n_1\;m_i s_{in},\\
    m_2^{(i)} &= -I^n_2\;m_{V^{(n)}}s_{in}+I^n_1\;m_i c_{in},\\
    m_3^{(i)} &= -I^n_2\;iM_W c_{in} +I^n_1\;i\frac{m_{V^{(n)}}m_i}{M_W}s_{in},\\
    m_4^{(i)} &= I^n_2\;iM_W s_{in}+I^n_1\;i\frac{m_{V^{(n)}}m_i}{M_W}c_{in},
  \end{aligned}
\end{equation}
where $m_i$ is identified as the mass of the zero-mode {\it up-type} fermion and $c_{in}=\cos(\alpha_{in})$ and $s_{in}=\sin(\alpha_{in})$ with $\alpha_{in}$ as defined earlier.

And the mass parameters $M_x^{(i,j)}$ are expressed in the following way \cite{Datta:2015aka}:
\begin{equation}\label{Mparameters}
  \begin{aligned}
    M_1^{(i,j)}  &= I^n_1\;m_j c_{in},\\
    M_2^{(i,j)}  &= I^n_1\;m_j s_{in},\\
    M_3^{(i,j)}  &= I^n_1\;i\frac{m_{V^{(n)}}m_j}{M_W}c_{in},\\
    M_4^{(i,j)}  &= I^n_1\;i\frac{m_{V^{(n)}}m_j}{M_W}s_{in},
  \end{aligned}
\end{equation}
where $m_j$ is identified as the mass of the zero-mode {\it down-type} fermion. 

In all the Feynman vertices the factors $I^n_1$ and $I^n_2$ are identified as the overlap integrals given in the following \cite{Datta:2015aka}

\begin{equation}
I^n_1 = 2\sqrt{\frac{1+\frac{r_V}{\pi R}}{1+\frac{r_f}{\pi R}}}\left[ \frac{1}{\sqrt{1 + \frac{r^2_f m^2_{f^{(n)}}}{4} + \frac{r_f}{\pi R}}}\right]\left[ \frac{1}{\sqrt{1 + \frac{r^2_V m^2_{V^{(n)}}}{4} + \frac{r_V}{\pi R}}}\right]\frac{m^2_{V^{(n)}}}{\left(m^2_{V^{(n)}} - m^2_{f^{(n)}}\right)}\frac{\left(r_{f} - r_{V}\right)}{\pi R},
\label{i1}
\end{equation}

\begin{equation}
I^n_2 = 2\sqrt{\frac{1+\frac{r_V}{\pi R}}{1+\frac{r_f}{\pi R}}}\left[ \frac{1}{\sqrt{1 + \frac{r^2_f m^2_{f^{(n)}}}{4} + \frac{r_f}{\pi R}}}\right]\left[ \frac{1}{\sqrt{1 + \frac{r^2_V m^2_{V^{(n)}}}{4} + \frac{r_V}{\pi R}}}\right]\frac{m_{V^{(n)}}m_{f^{(n)}}}{\left(m^2_{V^{(n)}} - m^2_{f^{(n)}}\right)}\frac{\left(r_{f} - r_{V}\right)}{\pi R}.
\label{i2}
\end{equation}
\end{appendices}


\providecommand{\href}[2]{#2}\begingroup\raggedright\endgroup

\end{document}